\documentclass[epj]{svjour}

\usepackage{color}
\usepackage{graphicx}

\usepackage{amssymb}
\usepackage{booktabs} % for nice tables
\usepackage{mathtools}

\title{Resolution of  hyper-triton chemical freeze-out puzzle in high energy nuclear collisions}

\author{O. V.  Vitiuk\inst{1, 2}, 
	K. A. Bugaev\inst{1, 3},
	E. S. Zherebtsova\inst{4, 5},
	D. B. Blaschke\inst{4, 6, 7},
	L. V.  Bravina\inst{2},
	E.  E.  Zabrodin\inst{2, 8}, 
	G. M. Zinovjev\inst{3}\\
}
\authorrunning{O. V.  Vitiuk}

\institute{                    
	\inst{1} Department of Physics, Taras Shevchenko National University of Kyiv, 03022 Kyiv, Ukraine\\
	\inst{2} University of Oslo, POB 1048 Blindern, N-0316 Oslo, Norway\\  	
	\inst{3} Bogolyubov Institute for Theoretical Physics, Metrologichna str. 14$^B$, Kyiv 03680, Ukraine\\
	\inst{4} National Research Nuclear University (MEPhI), Kashirskoe Shosse 31, 115409 Moscow, Russia\\
	\inst{5} Institute for Nuclear Research, Russian Academy of Science, 108840 Moscow, Russia\\
	\inst{6} Institute of Theoretical Physics, University of Wroclaw, Max Born Pl. 9, 50-204 Wroclaw, Poland\\
	\inst{7} Bogoliubov Laboratory of Theoretical Physics, JINR Dubna, Joliot-Curie Str. 6, 141980 Dubna, Russia\\
	\inst{8} Skobeltsyn Institute of Nuclear Physics, Moscow State University, 119899 Moscow, Russia
}


\abstract{The recently developed hadron resonance gas model with multicomponent hard-core repulsion is used to address and resolve the long standing problem to describe the light nuclear cluster multiplicities  including the hyper-triton measured by the STAR Collaboration, known as the hyper-triton chemical freeze-out puzzle.  An unprecedentedly accurate description is obtained for the hadronic and other light nuclear cluster data measured by STAR at the collision energy  $\sqrt{s_{NN}} =200$ GeV and by ALICE at $\sqrt{s_{NN}} =2.76$ TeV.   This success is achieved by applying the new strategy of analyzing the light nuclear cluster data and by using the value for the hard-core radius of the (anti-)$\Lambda$ hyperons found in earlier work. 
One of the most striking results  of the present  work is that for  the most probable scenario of  chemical freeze-out for the STAR energy the obtained parameters allow to simultaneously  reproduce the values of the experimental ratios $S_3$ and $\overline{S}_3$   which were not included in the fit.
}

\begin{document}
	
\maketitle

\section{Introduction} \label{Intro}
	
The  yields of light (anti-) and (hyper-) nuclei, i.e. the light nuclear clusters,   measured in heavy ion collisions at high energies 
triggered  a vivid discussion about the proper theoretical formulation of the cluster formation process \cite{Vitiyuk_Ref1,Vitiyuk_Ref2,Vitiyuk_Ref3}, 
since  at the moment there are serious conceptual questions related to it.
%more puzzles than the reliable answers. 
Such nuclei like the deuteron (d),  helium-3 ($^3$He),  helium-4 ($^4$He), hyper-triton ($^3_\Lambda$H) and their antiparticles 
should be extremely  sensitive to the assumptions of the model description, since their binding energies are much lower than the  typical hadronic 
mass scales and the chemical freeze-out (CFO)  temperature. 
Two major approaches with different underlying theoretical concepts, namely the coalescence models  and the versions of the thermal hadron resonance gas model (HRGM), have been developed to describe  the light nuclear cluster production in heavy-ion collisions, see, e.g., \cite{Andronic:2017pug,Braun-Munzinger:2018hat,Scheibl:1998tk,Steinheimer:2012tb,Oliinychenko:2018ugs,Coalesc1}.
They are applied to describe the recent data, but not all of  them are successfully reproduced \cite{Vitiyuk_Ref1,Vitiyuk_Ref2,Vitiyuk_Ref3,Coalesc1}.
In particular, the problematic  hyper-triton  ratios (PHTR)   $_\Lambda^3 \overline{H}/ ^3 \overline{He}$ and $_\Lambda^3 {H}/ ^3 {He}$ measured  by 
the STAR Collaboration at  $\sqrt{s_{NN}} =200$ GeV \cite{STARA1,STARA2}  are not explained up to now. 

%{\color{blue}
In this work we suggest a simple, but elegant solution of the puzzle of the PHTR  measured  by the STAR Collaboration at  $\sqrt{s_{NN}} =200$ GeV.
It allows us to simultaneously  describe  all hadronic and  light nuclear cluster multiplicities which were obtained at this collision energy previously and 
very recently \cite{STARA3}.
%}
A similar analysis is performed for the ALICE LHC data \cite{KAB_Ref1a,KAB_Ref1b,KAB_Ref1c} measured at $\sqrt{s_{NN}} =2.76$ TeV.
The newly developed HRGM  \cite{HRGM_IST1,HRGM_IST2,HRGM_IST3,HRGM_IST4,HRGM_IST5,HRGM_IST6} based on the induced surface tension 
(IST) equation of state (EoS) \cite{IST1,IST2,IST3,IST4b,IST4,IST5} allows us not only to achieve an unprecedentedly high quality of description of hadronic and nuclear cluster yields and/or their ratios, but it also provides a new and high standard for HRGM  existing nowadays. 

Using this opportunity we would like to analyze the reasons why the other versions of HRGM fail to resolve the PHTR puzzle and  to discuss what problems will face  the HRGM community once the accuracy of measured hadronic and the light nuclear cluster yields will be improved by an order of magnitude in the future experiments on NICA  JINR and FAIR GSI. 

{
In addition we continue to refine the novel strategy to analyze the light nuclear cluster data which was first  applied to the analysis of the ALICE data  \cite{KAB_Ref1a,KAB_Ref1b,KAB_Ref1c}  measured at 
$\sqrt{s_{NN}} =2.76$ TeV  in \cite{HRGM_IST5} and then it was successfully implemented  to the analysis of the STAR data  \cite{STARA1,STARA2,STARA3} in \cite{HRGM_IST6}. 
This strategy is based on the simultaneous use of two formulations of the IST EoS which, nevertheless, employ a different treatment of the hard-core repulsion. The first  approach is based on the classical excluded volumes of light nuclear clusters with all hadrons found in \cite{HRGM_IST5} and it is, indeed, the IST EoS.  The second, an approximative, but rather accurate  approach called the bag model radii (BMR) EoS \cite{HRGM_IST6}, employs an approximate treatment of the hard-core radii of light nuclear clusters in a pion dominated medium \cite{HRGM_IST5,HRGM_IST6}.  
By construction both approaches should give the  same  quality of data description for some common values of CFO  parameters. 
In this way one can remove the freedom in choosing the most probable hypothesis of CFO. 
Here we generalize such a strategy to arbitrary mixtures of hadrons with hard-core repulsion. 
As a result, the generalized strategy allows us to easily resolve the PHTR puzzle without spoiling the  perfect description of the  STAR and ALICE data 
mentioned above. 
Predictions for  the yields of popular nuclear clusters and for some exotic nuclei are also made. 

The work is organized as follows. 
In Sect. 2 we thoroughly discuss the necessary ingredients for the successful description of hadronic and light nuclear cluster yields and inspect the pitfalls of the other versions of HRGM. 
In addition in Sect. 2   the necessary mathematical apparatus is outlined. 
The analyses  of the STAR and ALICE data mentioned above is given in Sect. 3, while the conclusions are summarized in Sect. 4.

\section{Main elements of IST-based HRGM EoS}

\subsection{Justification of multicomponent hard-core repulsion}

The success of advanced versions of the HRGM  is based on a balance between  the simple parameterization of
the interaction among the hadrons and  the  detailed accounting for such properties of hadrons and hadronic resonances
as  their masses, charges, degeneracies, widths and decays into other hadrons. 
On the one hand, 
%{\color{blue} 
similarly  to the  hard-core repulsion among atoms \cite{Ebeling:2008mg,Ropke:2018ewt},
the one among hadrons reflects the fact that real hadrons are not point-like particles and the 
Pauli exclusion principle applied to the internal constituents of hadrons  generates a repulsive interaction  between them. 
In other words,  the quarks and anti-quarks, belonging to different hadrons,  generate a very strong repulsion due to the 
Pauli blocking effect \cite{Blaschke:2020}.  
Note that very recently the validity of this qualitative picture was explored quantitatively in an explicit calculation of the 
chirally improved quark Pauli blocking effect among nucleons as three-quark bound states in nuclear matter \cite{Blaschke:2020}
where also the corresponding nucleonic excluded volume parameters and hard-core radii were obtained.
%} {\color{red}  \large David, please check whether it is correct and improve it, if necessary!}
On the other hand, such a simple parameterization of hadronic interaction  is well justified by 
the strong cancellation  of the attractive and repulsive interaction \cite{Prakash1992} known from the  HRGM with  the 
quantum second virial coefficients of hadrons. Therefore, the residual  deviation  from the ideal gas with a weak repulsion
can be taken into account as the hard-core repulsion via the  classical second virial coefficients of hadrons. 

%Considering the HRGM with hard-core repulsion as a sufficiently realistic  EoS of  hadronic matter,  one can be sure that  its pressure will never  exceed the one of the quark-gluon plasma.
It is important to stress that a sufficiently realistic EoS of hadronic matter with hard-core repulsion prevents the pressure in HRGM from exceeding the one of the quark-gluon plasma.
The latter may happen, however,   if the hadronic matter  is  treated as a
mixture of ideal gases of  all known hadrons and hadronic resonances \cite{IST1,IST2,Satarov10,Satarov15}. 
Thus, the presence of hard-core repulsion provides an agreement with the simulations of  lattice quantum chromodynamics (QCD) 
\cite{KAB_lqcd1,KAB_lqcd2}.

Moreover,   an additional  reason to consider  the HRGM with hard-core repulsion as hadronic matter EoS
in the vicinity of CFO is a purely  practical one: 
since  the hard-core repulsion is a contact interaction, 
the energy per particle of such an EoS coincides with  the one of the ideal gas, 
even for the case of  quantum  statistics  \cite{IST2,IST3}. 
As a result, to model  the evolution of the system created in heavy ion collisions  between the 
CFO and  the kinetic freeze-out   one does not need to solve a  hard mathematical problem  \cite{KABkinFO1,KABkinFO2} to somehow 
transform  the potential energy of interacting hadrons  into their kinetic energy 
and into the masses of  particles which appear due to resonance decays. 

There are no a priori  arguments to believe that the hard-core radius of all hadronic species should be the same,
since the degree of  cancellation of repulsive and attractive forces depends on the pair of interacting hadrons \cite{Prakash1992}. 
Therefore, a multicomponent  HRGM (MHRGM), i.e. the HRGM with several different hard-core radii of  hadrons,  is required. 
The very high quality of  the data description with $\chi^2/dof  \simeq 1.15$ \cite{MHRGM1,MHRGM2,MHRGM3}  obtained by 
considering the  individual  hard-core radii $R_\pi$ for pions  and $R_K$ for kaons, i.e. just two extra  parameters,  in addition to the traditional HRGM \cite{PBM06} which employs the hard-core radii $R_b$ of baryons and  $R_m$ of other mesons, clearly demonstrates the advantages of  the MHRGM.  

Using the MHRGM  it was possible for the first time to correctly reproduce the peaks of  $K^+/\pi^+$ and $\Lambda/\pi^-$ ratios without spoiling the high quality fit 
of other hadron yield ratios  \cite{MHRGM1,MHRGM2}. 
Then, by adding into the MHRGM the hard-core repulsion of (anti-)$\Lambda$ hyperons with their own hard-core radius
it was possible to finally  resolve the anti-$\Lambda$ hyperon puzzle which in other HRGM like the one presented in \cite{PBM06} remained unsolved for years.  
Such an approach allowed us  to  get  a very accurate description of all independent  hadron multiplicity ratios measured in central nuclear collisions  at 
the collision energies $\sqrt{s_{NN}}  = 2.7,$ $ 3.3, 3.8, 4.3, 4.9, 6.3, 7.7, 8.8, 9.2, \,12.3, \, 17.3, 62.4, 130,$ \\ $ 200$ GeV   with $\chi^2/dof \simeq 0.96$ \cite{Sagun14}. 
Despite this great success  \cite{Sagun14,HRGM_IST1,HRGM_IST2,HRGM_IST3} achieved  in the  solution \cite{Sagun14} of the anti-$\Lambda$ puzzle  \cite{PBM06}, the present approach and the  value  $R_\Lambda = 0.085$ fm for the hard-core radius of (anti-)$\Lambda$ hyperons  \cite{HRGM_IST1,HRGM_IST2,HRGM_IST3} 
have not yet been used in other, far less realistic versions of the HRGM, and, moreover,  this value  was  criticized on several occasions.  
Therefore, in this work we will explicitly  demonstrate that the hyper-triton nuclear cluster and similar exotic light nuclei can provide us with an accurate tool to reliably  determine the individual hard-core radii of  short-lived hadrons like $\Lambda$ hyperons, if they form even a loosely bound state with nucleons. 
     
It is apparent that the reliable analysis of light  nuclear cluster multiplicities also requires an EoS with multicomponent hard-core repulsion \cite{HRGM_IST1}. 
However, until very recently  there were two  principal problems that did not allow one to develop the correct treatment of hard-core repulsion of  light  nuclear clusters. Firstly,   the classical second virial coefficients of   light  nuclear clusters with hadrons were unknown and, hence, it was unclear how to correctly  account for the hard-core repulsion among the light nuclei and hadrons in the MHRGM.  Secondly,   the Van der Waals treatment of multicomponent mixtures of gases is rather inconvenient, since  for $N$ different hard-core radii of particles it is necessary 
 to solve the  system of  (at least) $N$ transcendental equations and each of  these equations contains a few  hundreds of double integrals \cite{MHRGM1,MHRGM2,MHRGM3,Sagun14}.  As a result, both a further development of  the MHRGM based on Van der Waals   approximation and its application to  the cases  $N \gg 1$ have no perspective.  
 
Due to the fact that the classical second virial coefficients of  light nuclear clusters were unknown, oversimplified and, hence, unrealistic assumptions about  their interaction  with hadrons were used. 
As a result such versions of the HRGM were unable to reproduce the available data with high accuracy  and this fact stimulated  several research groups of the heavy ion physics community to concentrate on the problem of  light nuclear clusters and, in particular, on the PHTR. 
An almost complete list of the simpler versions of the HRGM can be found in \cite{Vitiyuk_Ref_Doenigus2020}. 
Here we discuss the typical oversimplifying assumptions.
Historically, the first version of HRGM that considered both hadrons and light nuclei clusters, treated them overly simplified as point-like particles \cite{KAB_Jean}.  
In the second simplified  HRGM description it was assumed that  the hard-core radii of all light nuclear clusters are equal to the hard-core radius $R_b$  of  baryons  \cite{KAB_Ref2}.
 
Surprisingly, these oversimplified HRGM of Refs. \cite{KAB_Jean,KAB_Ref2} are able to  describe the  experimental data for
hadronic and light nuclear clusters to some extent.  
A close inspection shows \cite{HRGM_IST1,HRGM_IST2,HRGM_IST3} 
that the reason for such an apparent success, which  turns out to be  a ``pyrrhic victory``,  is that the local minimum of
the  light nuclear clusters $\chi^2_A$ as a function of their  CFO temperature $T_A$,  CFO baryonic chemical potential $\mu_B$ 
and  CFO volume $V_A$  is very shallow and wide. Hence, the narrow  $\chi^2_h$ minimum generated by the hadronic data is always dominant and, hence,  
one can poorly reproduce the total set of the hadronic and  light nuclear clusters experimental data.  
Apparently, this is a self-cheating approach, since several erroneous assumptions can, in principle, produce the results which are  not far from the truth on which, however,  one cannot rely.  
Therefore,  both from the academic and from the practical points of view it is absolutely necessary 
to develop a more realistic EoS to treat the  light nuclear clusters.
 
Note that the existing shallow and wide  minimum of $\chi^2_A$ of   light nuclear clusters  creates the  problem  
to  determine  the most reliable  CFO parameters of light nuclear clusters in the case, if their CFO occurs separately from that of the hadrons \cite{HRGM_IST3,HRGM_IST4,HRGM_IST5}.  
Such a problem cannot be resolved easily, even if one employs the most advanced MHRGM based on the IST EoS for the mixture of hadrons and light nuclear clusters \cite{HRGM_IST2,HRGM_IST3}.  
Therefore, we were forced to search for an entirely new approach \cite{HRGM_IST5,HRGM_IST6} that is presented below.
 
\subsection{New strategy to analyze light nuclear cluster data}

In this work  we are using the most  advanced MHRGM with the IST EoS to fit the  data measured by STAR and ALICE. 
This MHRGM was first heuristically obtained in \cite{HRGM_IST5}, whereas in Ref.  \cite{HRGM_IST6} it was rigorously derived 
from the grand canonical partition using the self-consistent method to treat the excluded volumes of multicomponent mixtures
of classical \cite{IST5} and quantum \cite{IST4} particles with hard-core repulsion.
 In general the MHRGM with the IST EoS is a system of coupled equations for the  pressure $p$ of considered system  and its  surface 
 tension coefficient $\Sigma$ that is 
  induced by the hard-core repulsion.  This system   can be written in the  form 
\begin{equation}
	\label{eq_1}
	\begin{dcases}
	&\hspace*{-2.77mm}p \equiv \sum\limits_{k \in h, A} p_k = T \hspace*{-0.77mm} \sum\limits_{k \in h, A} \hspace*{-0.77mm}
	 \phi_k \exp \left[\frac{\mu_k - p V_k -\Sigma S_k}{T} \right], \\
&\hspace*{-2.77mm}\Sigma \equiv  \hspace*{-0.77mm} \sum\limits_{k \in h, A} \hspace*{-0.77mm} \Sigma_k = \hspace*{-0.77mm} \sum\limits_{k  \in h, A} \hspace*{-0.77mm} R_k p_k \exp \left[-\frac{(\alpha-1) \Sigma S_k}{T} \right],\\
&\hspace*{-1.77mm}\mu_k = \mu_B  {\cal B}_k + \mu_S {\cal S}_k + \mu_{I_3}{\cal  I}_{3k}.
	\end{dcases}
\end{equation}
Here $p_k$ is the partial pressure of the k-th sort of particles, $\Sigma_k$ is their induced surface tension coefficient, 
$\mu_B$, $\mu_S$ and $\mu_{I3}$ are, respectively, the  baryonic, the strange and the third projection of isospin chemical potentials, while ${\cal B}_k$, ${\cal S}_k$, ${\cal I}_{3k}$,  respectively, denote the corresponding charges. 
$R_k$, $S_k$ and $V_k$ denote, respectively,  the  hard-core radius, the eigensurface and eigenvolume of the species $k$ of particles which are specified below. 
$\phi_k$ is the one-particle thermal partial density of the particle species $k$  with the Breit-Wigner mass distribution function 
\begin{equation}
	\label{eq_2}
	\begin{multlined}
	\phi_k = g_k  \gamma_S^{|s_k|} \int\limits_{M_k^{Th}}^\infty \frac{ d m}{N_k (M_k^{Th})}\frac{\Gamma_k}{(m-m_{k})^{2}+\Gamma^{2}_{k}/4} \times \\
	\times \int \frac{d^3 p}{ (2 \pi \hbar)^3 } \exp \left[{\textstyle  - \frac{ \sqrt{p^2 + m^2} }{T} }\right],
	\end{multlined}
\end{equation}
where $g_k$ is the degeneracy factor of the k-sort of hadrons, $\gamma_S$ is the strangeness suppression factor \cite{Rafelski}, $|s_k|$ is the number of valence strange quarks and antiquarks in this kind of hadrons. 
The quantity  $N_k(M^{Th}_k)$ denotes a corresponding normalization factor,  while $M^{Th}_k$ corresponds to the decay threshold mass of the hadron $k$. 
For stable particles and for light nuclear clusters one should take the limit $\Gamma_k \rightarrow +0$ in Eq.~(\ref{eq_2}).

The parameter $\alpha=1.25$ in Eq.~(\ref{eq_1}) allows us to reproduce not only the second, but also the third and the fourth virial coefficients of 
classical hard spheres \cite{HRGM_IST2,HRGM_IST3,IST2,IST3,IST4} and, hence, it enables us  to go beyond the Van der Waals approximation.
At the same time by  choosing $\alpha=1$ one obtains a convenient system of equations for the Van der Waals EoS with the
multicomponent hard-core repulsion \cite{HRGM_IST2,HRGM_IST3,IST2,IST3,IST4b,IST4}.

The summations in system \eqref{eq_1} are made over all sorts of hadrons $h$ and nuclei $A \in \{ 1, 2, 3, 4\}$ and their corresponding antiparticles which are considered as independent species. This system of equations is supplemented by the strange charge conservation law (see Ref. \cite{HRGM_IST2} for details). 

From the system \eqref{eq_1} one can determine  the thermal particle number density of k-sort of  particles 
\cite{HRGM_IST6} as
\begin{equation}
\label{eq_3}
\rho_k \equiv \frac{\partial  p}{\partial \mu_k} = \frac{1}{T} \cdot \frac{p_k a_{22} 
	- \Sigma_k a_{12}}{a_{11}a_{22} - a_{12}a_{21} }~,
\end{equation}
where the coefficients $a_{ij}$ are  defined as \cite{HRGM_IST6}
\begin{eqnarray}
\label{eq_4}
%\begin{multlined}
a_{11} &=& 1 + \sum_{k \in h, A} V_k \frac{p_k}{T}, \\ %\quad 
a_{12} &=& \sum_{k \in h, A} S_k \frac{p_k}{T}, \\
a_{21} &=& \sum_{k \in h, A} V_k\frac{\Sigma_k}{T}, \\% \quad
a_{22} &=& 1 + \alpha \sum_{k  \in h, A} S_k \frac{\Sigma_k}{T}       .     
%\end{multlined}
\end{eqnarray}
To fit the experimental yields of hadrons, to the thermal particle number density (\ref{eq_3})
 one has also to add the contributions coming from the decays of resonances. 
 For the known branching ratios $Br_{l\rightarrow k}$ of hadronic decays $l\rightarrow k$ one can write the total
yield of hadron $k$ as 
\begin{equation}
\label{eq_5}
N^{\rm tot}_k = V\left[ \rho_k+\sum_{l\neq k}\rho_l\, Br_{l\rightarrow k} \right],
\end{equation}
with the CFO volume $V$. 
Then, the ratio ${\cal R}_{lk} = {N^{\rm tot}_k }/{N^{\rm tot}_l}$ of yields of the hadrons $l$ and $k$ can be found. 

In the present  work we are using a combined fit of  particle  yields and ratios, as dictated by the available data and by numerical convenience. 
Hence, the total $\chi^2_{\rm tot}(V)$ is
\begin{eqnarray}
\label{eq_6}
%\begin{multlined}
\chi^2_{\rm tot}(V) &=& \chi^2_{R} + \chi^2_{Y}(V) \nonumber \\
&=& \hspace{-2mm} \sum_{ {k \neq l} \in R} \left[ \frac{{\cal R}_{kl}^{\rm theo} - {\cal R}_{kl}^{\rm exp}}{\delta {\cal R}_{kl}^{\rm exp}}\right]^2  
% \nonumber \\ &&
\hspace{-2mm} + \hspace{-1mm}\sum_{k \in Y} \left[ \frac{\rho_k(T) V - N^{\rm exp}_k}{\delta N^{\rm exp}_k}\right]^2\hspace{-2mm}, \nonumber \\
%\end{multlined}
\end{eqnarray}
where $\chi^2_R$ and $\chi^2_Y$ denote, respectively, the mean squared deviation for the ratios and for the yields.

All hard-core radii of hadrons, i.e. the hard-core radius of pions $R_\pi = 0.15$ fm, of  kaons  $R_K=0.395$ fm,
of  \mbox{(anti-)$\Lambda$} hyperons $R_\Lambda = 0.085$ fm, of other 
 baryons $R_b =0.365$ fm and the one of other mesons  $R_m=0.42$ fm, are taken from \cite{HRGM_IST2,HRGM_IST3} (new radii in terms of Refs. \cite{HRGM_IST2,HRGM_IST3}).  
 These radii  provide an excellent  description of all independent  hadron multiplicity ratios measured in central nuclear collisions at all AGS and all SPS energies,  at the RHIC highest energies  $\sqrt{s_{NN}} =62.4, 130$ and $200$ GeV and at the ALICE energy $\sqrt{s_{NN}} =2.76$ TeV  with 
 $\chi^2/dof \simeq 1.13$ \cite{HRGM_IST3}. 
 Then the eigensurface and eigenvolume of the $k$-th sort of hadrons  are defined as $S_k = 4 \pi R_k^2$  and $V_k = \frac{4}{3}\pi R_k^3$, respectively.
 
To take into account the classical excluded volumes of light nuclear clusters and hadrons  we employ  two approaches worked out in  \cite{HRGM_IST4,HRGM_IST5,HRGM_IST6} with one exception, namely here we consider the hyper-triton (HTR) differently compared to our previous studies in Refs. \cite{HRGM_IST4,HRGM_IST5,HRGM_IST6}.
Both of  these approaches  use  the  classical excluded volumes of light nuclear cluster of  $A \in \{1, 2, 3, 4\}$ baryonic constituents   and hadron $h$ 
\cite{HRGM_IST5,HRGM_IST6}
\begin{equation}\label{Eq7}
 b_{Ah} = b_{hA} =  A \frac{2}{3}\pi (R_b+R_h)^3\,, 
\end{equation}
which can be easily  found from the fact that all  light nuclear clusters analyzed here are roomy clusters.
An analysis of mean classical distances among the baryons inside of such clusters \cite{HRGM_IST6}  shows  one  that
it is possible to  freely translate the hadron $h$ with the hard-core radius $R_h$ around each  constituent of a nucleus  without touching  any other constituent  of   this  nucleus. 

The first of these approaches, named as  the IST EoS,   is rigorously derived using a self-consistent treatment of classical excluded volumes of light (anti-)nuclei and hadrons \cite{HRGM_IST6}. 
In this approach the  parameters $R_A$, $S_A$ and $V_A$ of light nuclear clusters with the mass number $A$ entering  the system (\ref{eq_1})  are given 
as functions of the baryon radius by
\begin{equation}
	\label{eq_8}
	R_A = A\cdot   R_b, ~ S_A = A \cdot  4 \pi R_b^2, ~ V_A = A \cdot  \frac{4}{3}\pi R_b^3 .
\end{equation}

The second approach is approximate and complementary to the first one. 
It is based on an approximate, but rather  accurate  treatment of the equivalent hard-core radius of roomy nuclear cluster and pions which are the dominating 
component of the HRG at the energy range of our interest. 
In this approach one can find an effective hard-core radius of nuclei as $R_A \simeq A^{1/3}R_b$, since  
the  hard-core radius of pions is very small and it generates a negligible correction to $R_A$  \cite{HRGM_IST6}. 
Consequently,  the eigensurface $S_A$  and eigenvolume $V_A $ of  light nuclear clusters can be found as 
\begin{equation}
	\label{eq_9}
R_A \simeq A^\frac{1}{3}R_b, ~\Rightarrow~ S_A \simeq A^\frac{2}{3}  \cdot  4 \pi R_b^2, ~~ V_A \simeq A \cdot  \frac{4}{3}\pi R_b^3. 
\end{equation}
The hard-core radius of light nuclear clusters defined in this way  is similar to the expression of the Bag Model  \cite{BMR}. 
Hereafter this model is called the BMR EoS.

Applying the Eqs. (\ref{eq_8}) and (\ref{eq_9})  to the  PHTR   $_\Lambda^3 \overline{H}/ ^3 \overline{He}$ and $_\Lambda^3 {H}/ ^3 {He}$ 
measured  by STAR at  $\sqrt{s_{NN}} =200$ GeV \cite{STARA1,STARA2}, we obtained a very good  overall  fit of  seven data points  for the CFO 
temperature of nuclei  $T_A =167.3 \pm 3.93$ MeV with   $\chi^2_A \simeq 9.6$ and for  $T_A =240.3 \pm 21.38$ MeV  with  $\chi^2_A \simeq 4.92$ 
(see  Fig. 4 and Table 3 in Ref. \cite{HRGM_IST6}). 
Nevertheless,  to our surprise,  even at the high temperature $T_A =240.3$ MeV the mean squared deviation 
of these PHTR alone  is large and it  defines the whole value of  $\chi^2_A \simeq 4.92$. 
In other words, even these advanced approaches are not doing a better job for the PHTR  than  the other thermal models cited in  \cite{Vitiyuk_Ref_Doenigus2020}. 

A close inspection of Eqs.~(\ref{eq_8}) and (\ref{eq_9})  enabled us to find an elegant solution to this puzzle, which, nevertheless,  requires  the most natural  generalization of the two approaches mentioned above to  the case of nuclei with $A$ constituents which can be divided into   $N_{s}$  different sorts 
\begin{equation}
	\label{eq_10}
	A = \sum_{k=1}^{N_s} n_k,  ~~ \text{with}~~ n_k \ge 1, 
\end{equation}
where $n_k \in Z$  is the number of  constituents  of the $k$-th sort  inside the nuclei.

For the IST EoS  approach the generalization (\ref{eq_10})   leads to  a set of simple replacements
\begin{equation}
\label{eq_11}
R_A \rightarrow \sum_{k=1}^{N_s} n_k R_k, ~ S_A \rightarrow \sum_{k=1}^{N_s} n_k S_k, ~ V_A \rightarrow \sum_{k=1}^{N_s} n_k V_k,
\end{equation}
where the sums are running over the constituent hadrons of sort $k$ having the hard-core radius $R_k$, the eigensurface 
$S_k$ and eigenvolume $V_k$. For the HTR  one can write the Eqs. (\ref{eq_11}) explicitly as
\begin{eqnarray}
\label{eq_12}
R_{HTR} &=& 2 \cdot R_b + R_{\Lambda} \nonumber, \\
S_{HTR} &=& 2 \cdot  4 \pi R_b^2 + 4 \pi R^2_{\Lambda}, \\
V_{HTR} &=&  2 \cdot \frac{4}{3}\pi R_b^3 + \frac{4}{3}\pi R^3_{\Lambda} \nonumber.
\end{eqnarray}
The resulting EoS is called hereafter the IST$\Lambda$ EoS. 

In order to modify the BMR EoS one should introduce the equivalent hard-core radius $R_{Ah}$ for a pair of a nucleus
with  $A$ constituents and an effective hadron $h$ with the mean hard-core radius $\overline{R}$ defined later. 
Then equating the excluded volume $\frac{2}{3}\pi(R_{Ah})^3$ to the second virial coefficient  of such a pair  $b_{Ah}$ of Eq.  (\ref{Eq7}), one finds 
\begin{equation}
	\label{eq_13}
	b_{Ah} = \frac{2}{3}\pi\left(R_A + \overline{R}\right)^3 = \frac{2}{3}\pi\left[ \sum_{k=1}^{N_s} n_k \left(R_k + \overline{R}\right)^3 \right],
\end{equation}
or explicitly 
\begin{equation}
\label{eq_14}
R_{Ah} = R_A + \overline{R} = \left[ \sum_{k=1}^{N_s} n_k \left(R_k + \overline{R}\right)^3 \right]^\frac{1}{3} .
\end{equation}
Now 
from the expression for $R_{Ah}$  one can determine the effective hard-core radius of a nucleus in a hadronic medium
with the mean hard-core radius  $\overline{R}$
\begin{equation}
\label{eq_15}
R_A = R_{Ah} - \overline{R} = \left[ \sum_{k=1}^{N_s} n_k \left(R_k + \overline{R}\right)^3 \right]^\frac{1}{3} - \overline{R}.
\end{equation}
Applying Eq.  (\ref{eq_13}) to the HTR  in a medium dominated by pions, one obtains
\begin{equation}
\label{eq_16}
R_{HTR} = \left[2 \left(R_b + R_{\pi}\right)^3 + \left(R_{\Lambda} + R_{\pi}\right)^3 \right]^\frac{1}{3} - R_{\pi}.
\end{equation}
Hereafter  this prescription for the hard-core radius of light nuclear clusters is  called as the BMR$\Lambda$ EoS.

Note that due to the inequality  $R_{\Lambda} \ll R_b$ the  terms with $R_{\Lambda}$ are almost negligible and, hence,  the excluded volume of  HTR is practically the same as  the deuteron one.

If the medium has different properties, for instance, if  it is baryon dominated, the mean hard-core radius of  the hadronic mixture can be found from the solution of  system (\ref{eq_1})  as
\begin{equation}
	\label{eq_17}
	\overline{R} =  \frac{\Sigma}{p} \simeq \frac{\sum\limits_{k  \in h} \hspace*{-0.77mm} R_k p_k \exp \left[-\frac{(\alpha-1) \Sigma S_k}{T} \right]}{\sum\limits_{k \in h} p_k},
\end{equation}
where the contribution of nuclei in the sums above is neglected.  Note that the quantity $\overline{R}$, indeed, is defined
in this way in the self-consistent treatment of second virial coefficients of mixture of hadrons and nuclei developed recently  in  \cite{HRGM_IST6}. Eq.  (\ref{eq_17}) shows that the main contribution to the mean radius $\overline{R}$ is generated by the particles with highest value of partial pressure. Apparently, these are the lightest particles with the smallest hard-core radius.
For high temperatures and low baryonic charge densities these are pions, while for baryon rich matter it should be found from Eq.  (\ref{eq_17}) using the solution of system   (\ref{eq_1}).  
From Eq.  (\ref{eq_17}) one can see that 
for very  high particle number densities (typically above the CFO ones) the mean radius $\overline{R}$ gets smaller  due to the  suppression factors  $\exp \left[-\frac{(\alpha-1) \Sigma S_k}{T} \right] < 1$  
which are generated by the induced surface tension coefficient  for  $\alpha > 1$.  
}

\section{Analysis of light nuclei multiplicities measured in A+A collisions} \label{Sect3}

{ 
Following our original idea \cite{HRGM_IST4}, in this section we consider not only the traditional 
scenario of a simultaneous CFO of hadrons and light nuclear clusters (single CFO), but also the 
scenario of their  separate freeze-outs (separate CFO).  In contrast to hadrons which are tightly bound
objects the light nuclei  are loosely bound and, therefore, it is hard to believe that their evolution 
is similar to the one of hadrons. Consequently, there is no a priori reason to assume  that the CFO 
of  light nuclear clusters coincides with the one of hadrons. 

Nevertheless, the success of the MHRGM in  describing the multiplicities of hadrons and light nuclear clusters
provides an 
evidence that their thermalization mechanisms may be similar. In  our previous works \cite{HRGM_IST4,HRGM_IST5,HRGM_IST6}
we argued that the light nuclear clusters are produced at the moment of hadronization of quark-gluon bags.
Since the quark-gluon bags with  the Hagedorn mass spectrum \cite{KAB_Ref27} are perfect thermostats and perfect particle reservoirs  \cite{Thermostat1}, 
any particle or cluster emitted  by  such  a  bag will be produced   in  a full chemical and thermal equilibrium with it  \cite{Thermostat1}. 
In principle,  such a hypothesis is not only able to  explain  the fact that the light nuclear clusters appear in full chemical and thermal  equilibrium, but also   the scale  of  their  CFO temperatures extracted  from the ALICE  and STAR data in  \cite{HRGM_IST6}. Of course, the corresponding model is highly required, but before its development  one has to reliably extract the parameters of the CFO of light nuclear clusters. 

It is also necessary to remind that the hypothesis of simultaneous  chemical and kinetic freeze-out of heavy particles \cite{KAB_EarlyFO1,KAB_EarlyFO2,KAB_EarlyFO3,KAB_EarlyFO4} which do not form resonances with pions was suggested a long time ago. 
It was argued \cite{KAB_EarlyFO1,KAB_EarlyFO2,KAB_EarlyFO3,KAB_EarlyFO4}  that such a simultaneous freeze-out  occurs at the hadronization of quark-gluon bags.
The validity of  such a hypothesis was for the first time demonstrated for  $\Omega$ hyperons and $J/\psi$ and $\psi^{\prime}$ mesons in Refs. \cite{KAB_EarlyFO1,KAB_EarlyFO2} and recently it was  extended  \cite{HRGM_IST6} to the CFO of  light nuclear clusters 
in high energy nuclear collisions.

Besides, the hypothesis of  separate CFO was  proved to be very successful  in explaining the nuances of the CFO 
of strange hadrons  \cite{MHRGM3,SFO_ind}. The strangeness suppression factor 
$\gamma_S$ was  introduced in Ref. \cite{Rafelski} to quantify the deviation of strange charge from chemical equilibrium.
However, as it was for the first time shown in  \cite{MHRGM3,SFO_ind}  such a factor is unnecessary to describe the hadronic multiplicities, since the data  
can be perfectly  explained by the fact that the CFO of  strange hadrons occurs on a separate hyper-surface compared to the one for hadrons  which consist  of the  
$u$ and $d$ (anti)quarks \cite{MHRGM3,SFO_ind}.  
More sophisticated scenarios of the CFO of strange hadrons  were suggested  in Refs. \cite{SFO_ind2,SFO2}. 

The new strategy based on a combined description of the same set of data achieved  independently by  the IST and BMR EoS worked very successfully  \cite{HRGM_IST6} with the inaccurate prescription of Eq.  (\ref{eq_8}) for the HTR excluded volume  and the one of  Eq.  (\ref{eq_9}) for the effective hard-core radius of  HTR. It allowed us to easily distinguish the scenarios of single and separate CFO of light nuclear clusters. Therefore, it is a very intriguing question what this strategy will give us for the correct prescriptions  (\ref{eq_12}) and (\ref{eq_16}) worked out  here. 
}

\begin{figure}[th]
	\centering
	\includegraphics[width=\columnwidth]{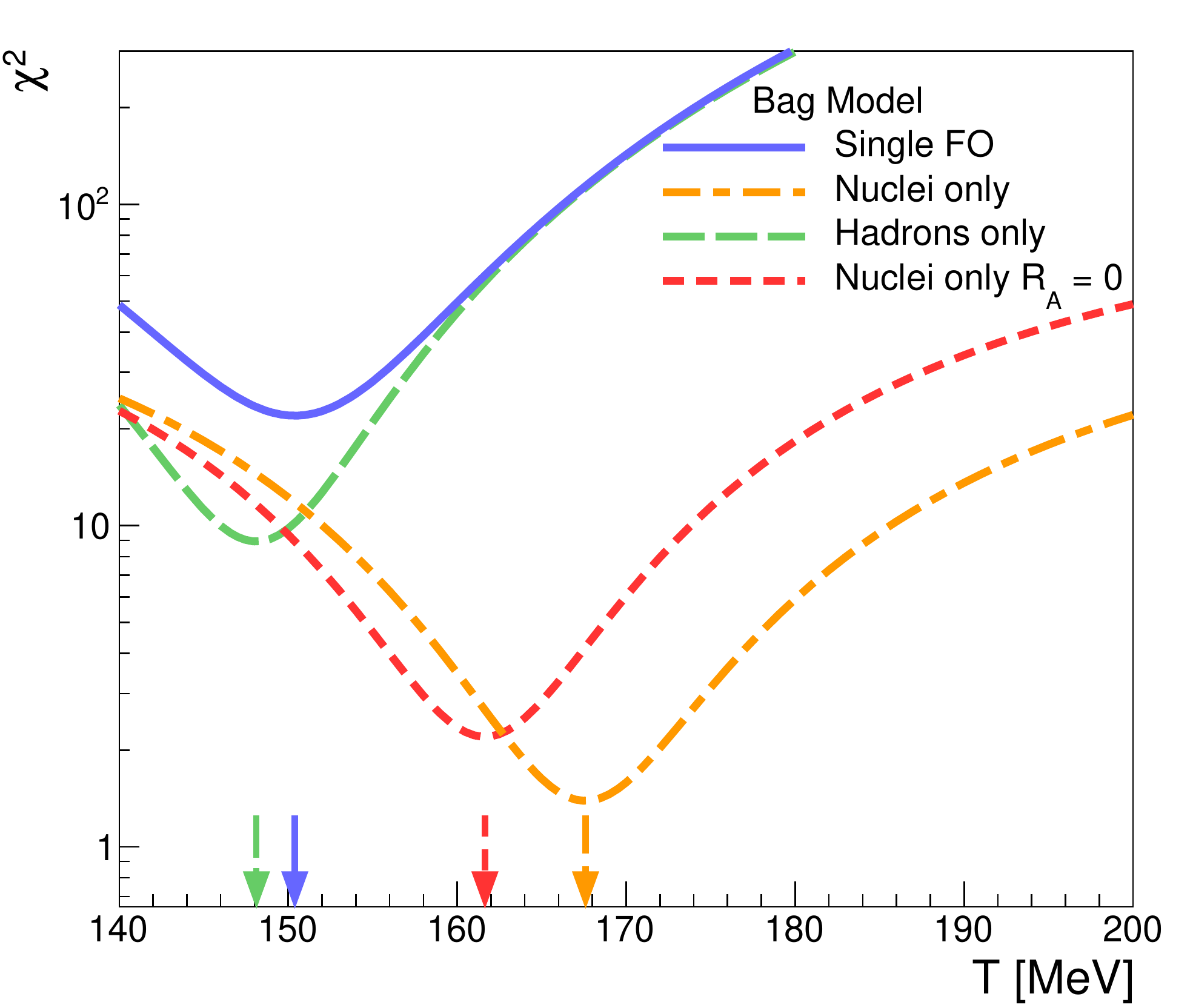}
	%\vspace*{-1mm}
	\includegraphics[width=\columnwidth]{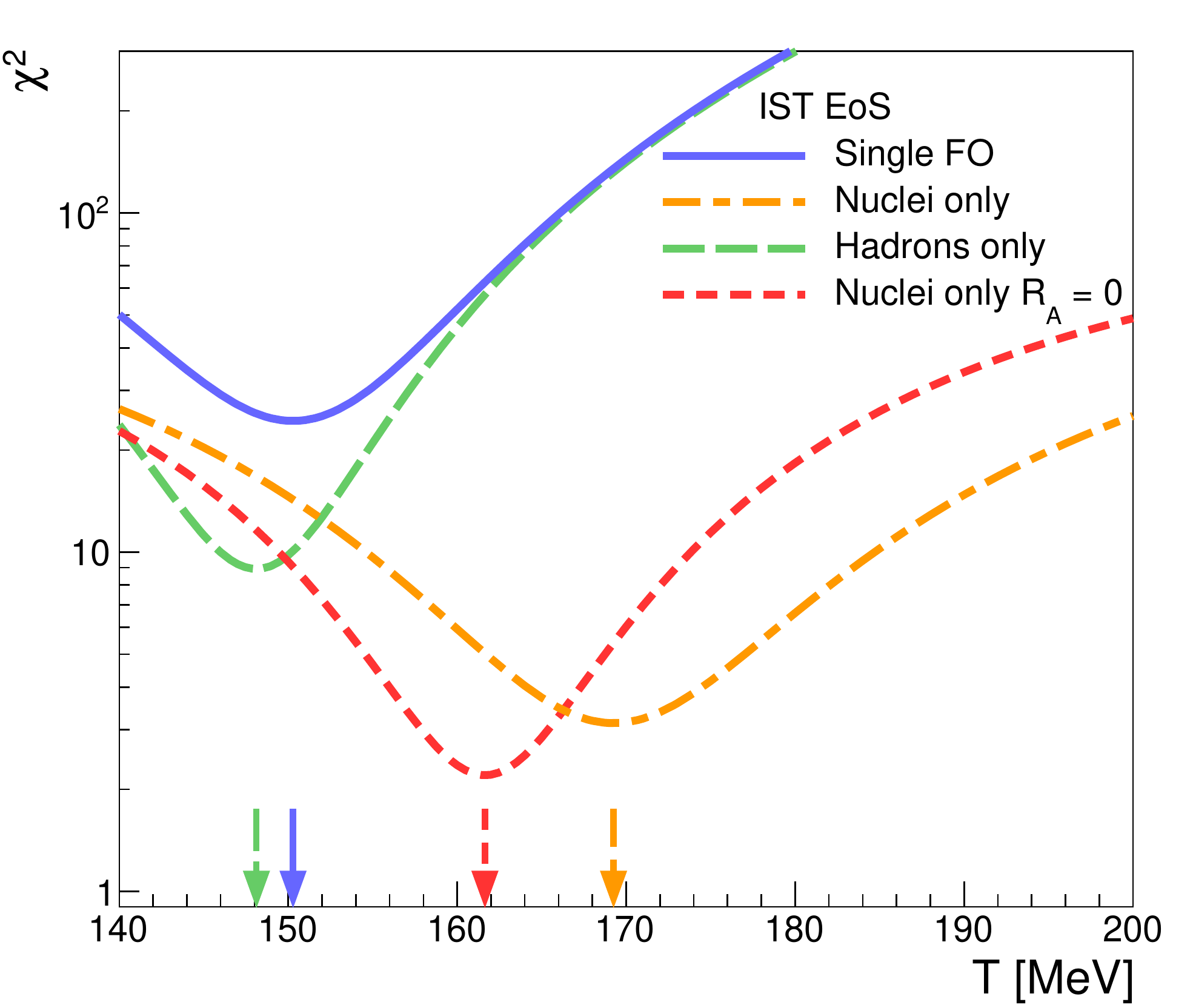}
	%\vspace*{-2mm}
	\caption{Temperature dependence of $\chi^2_{tot}$, $\chi^2_{h}$ and $\chi^2_{A}$ for fit of ALICE data measured at $\sqrt{s_{NN}} = 2.76$ TeV.  
		{\bf Upper panel:} The dependence obtained with the BMR$\Lambda$ EoS.
		{\bf Lower panel:} The dependence obtained with the IST$\Lambda$ EoS.}
	\label{ALICE_chi2}
\end{figure}

\begin{figure}[ht]
	\centering
	\includegraphics[width=\columnwidth]{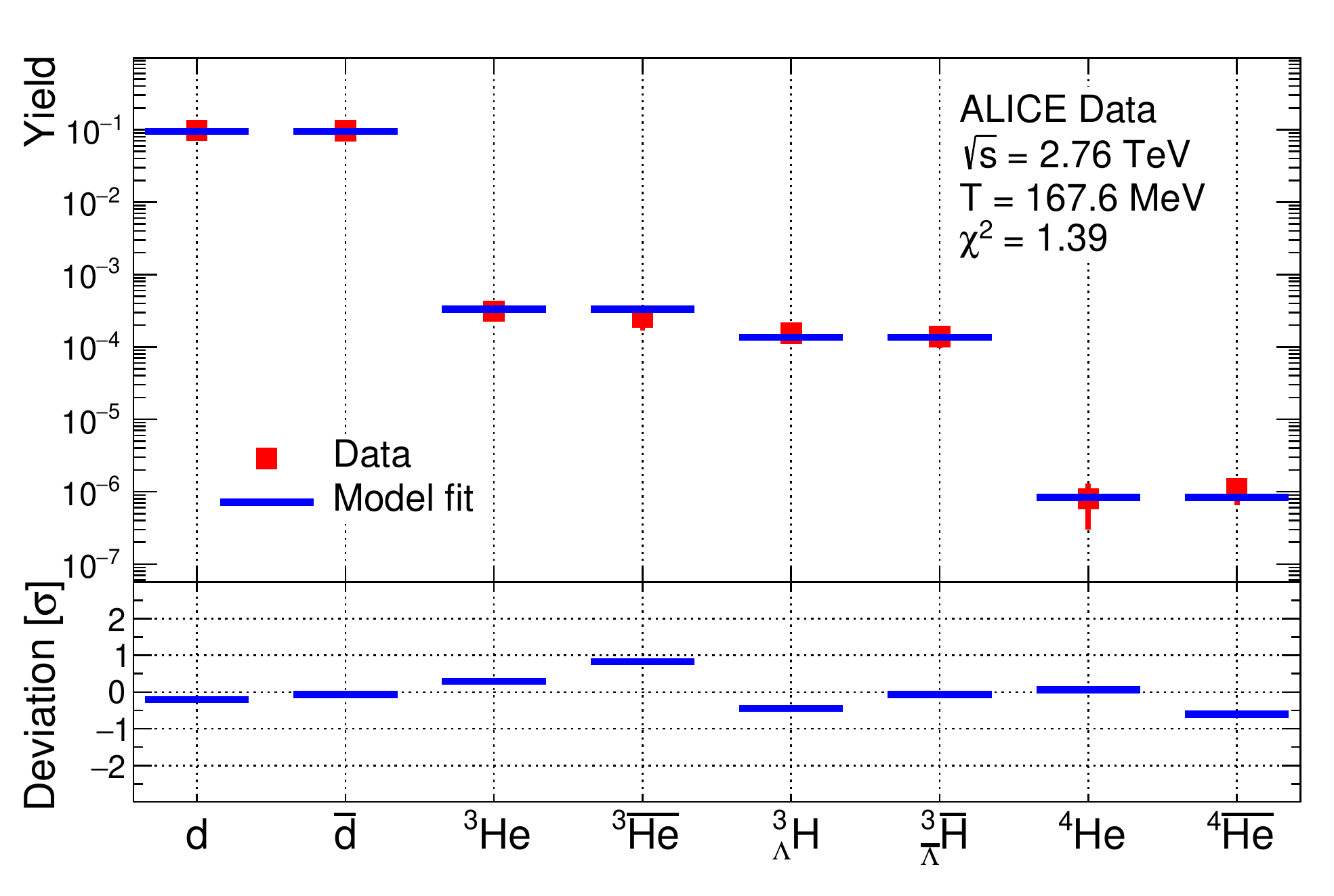}
	%\vspace*{-1mm}
	\includegraphics[width=\columnwidth]{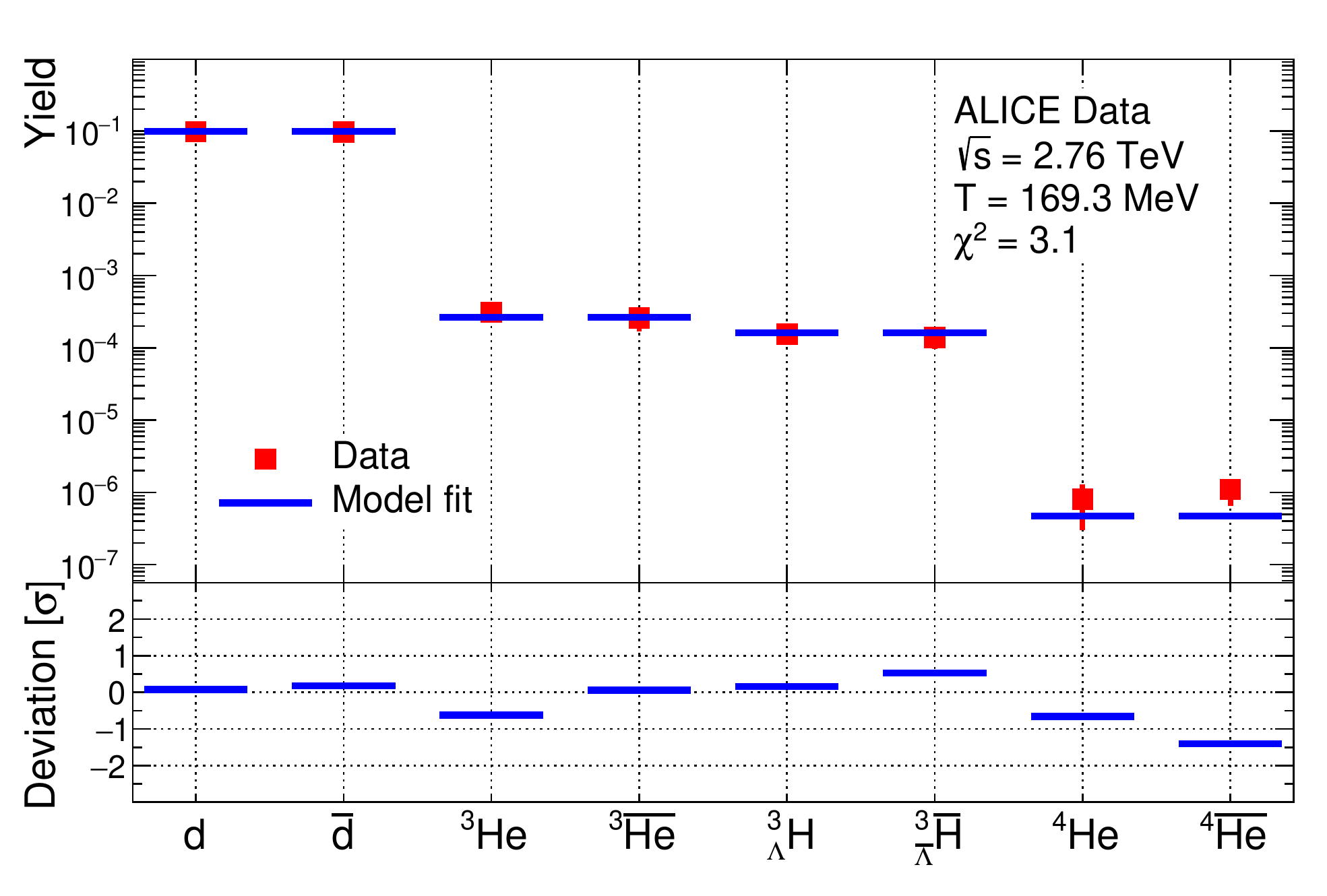}
	%\vspace*{-2mm}
	\caption{The yields of nuclear clusters measured at $\sqrt{s_{NN}} = 2.76$ TeV by ALICE vs. theoretical description in the scenario of separate CFO of light (anti)nuclei. Insertion shows the deviation of theory from data in the units of experimental error. 
		{\bf Upper panel:} The $\min\chi^2_A (V)$ corresponds to the BMR$\Lambda$ EoS.
		{\bf Lower panel:} Same as in the upper panel, but for the IST$\Lambda$ EoS. }
	\label{ALICE_fit}
\end{figure}

\begin{figure}[th]
	\centering
	\includegraphics[width=\columnwidth]{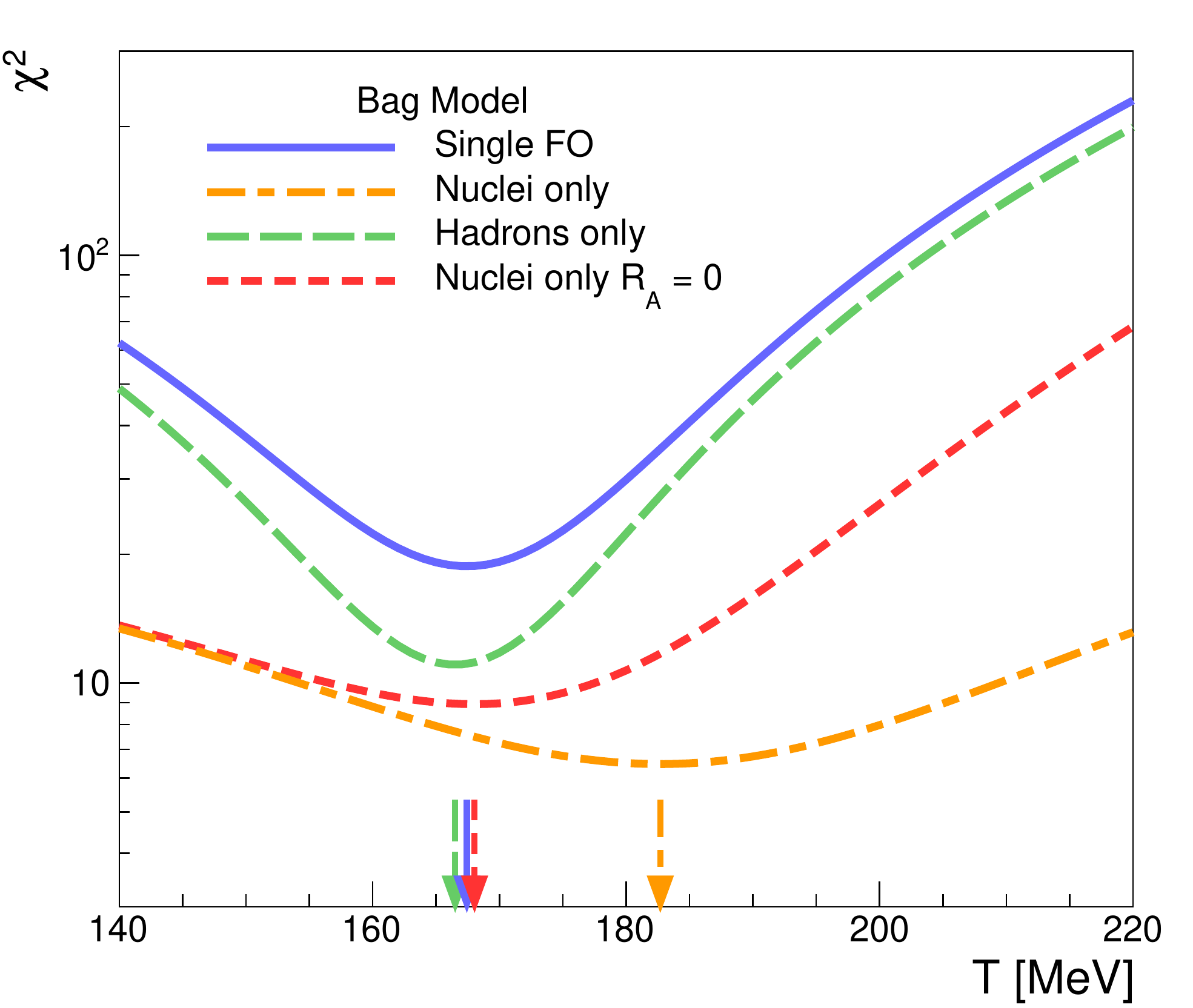}
	%\vspace*{-1mm}
	\includegraphics[width=\columnwidth]{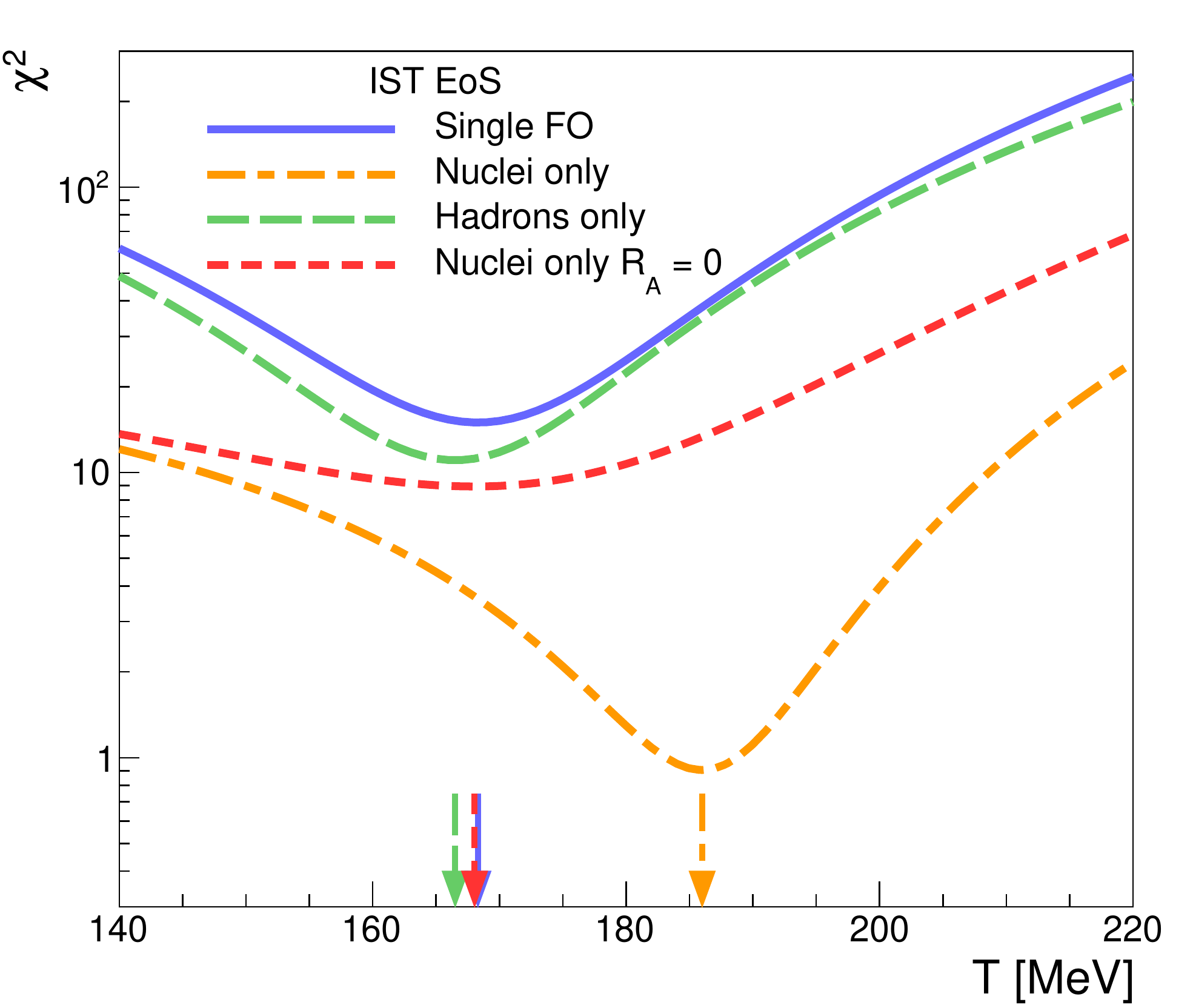}
	%\vspace*{-2mm}
	\caption{Temperature dependence of $\chi^2_{tot}$, $\chi^2_{h}$ and $\chi^2_{A}$ for fit of STAR data measured at $\sqrt{s_{NN}} = 200$ GeV.  
	{\bf Upper panel:} The dependence obtained with the BMR$\Lambda$ EoS.
	{\bf Lower panel:} The dependence obtained with the IST$\Lambda$ EoS.}
	\label{STAR_chi2}
\end{figure}

\begin{figure}[th]
	\centering
	\includegraphics[width=\columnwidth]{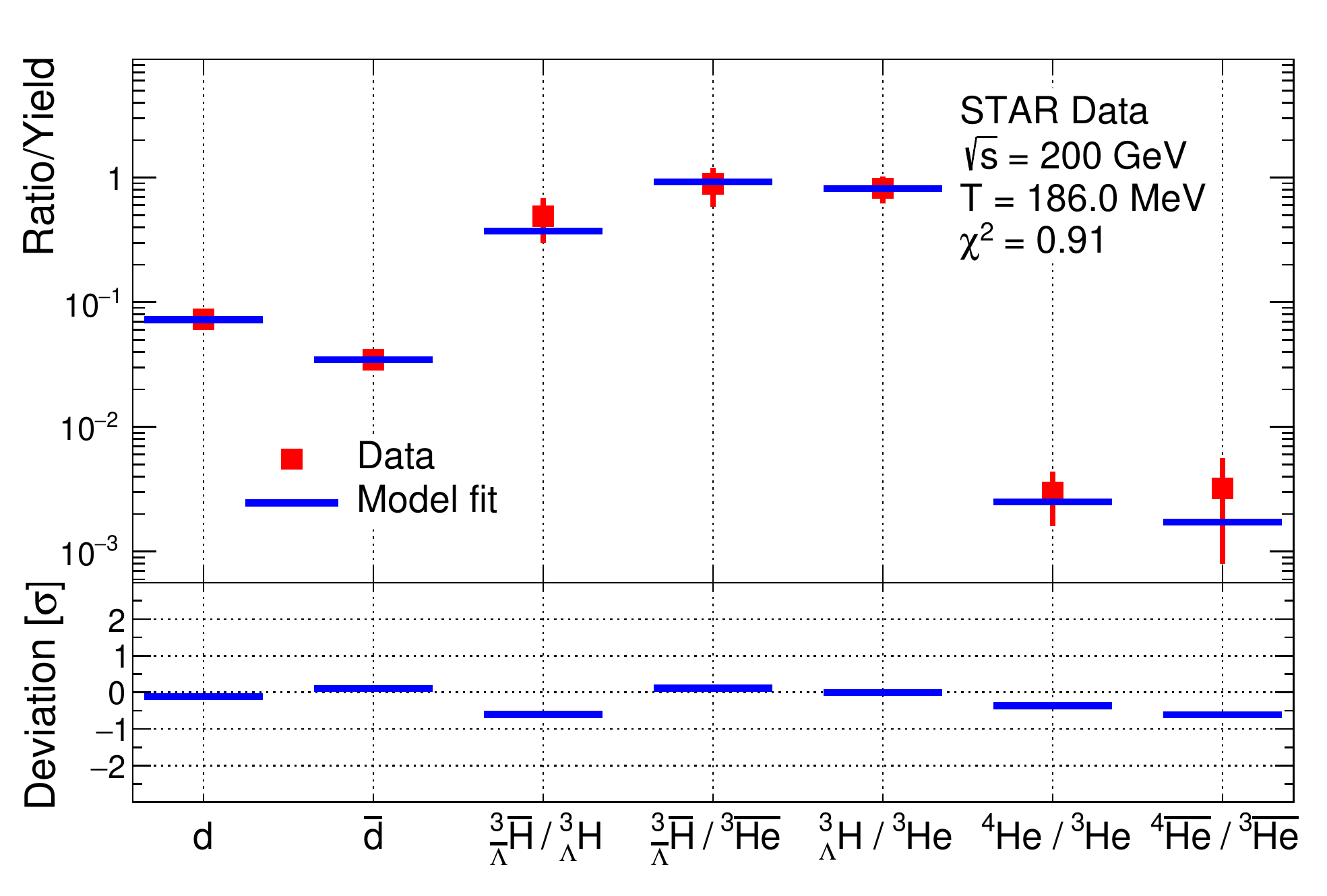}
	%\vspace*{-1mm}
	\includegraphics[width=\columnwidth]{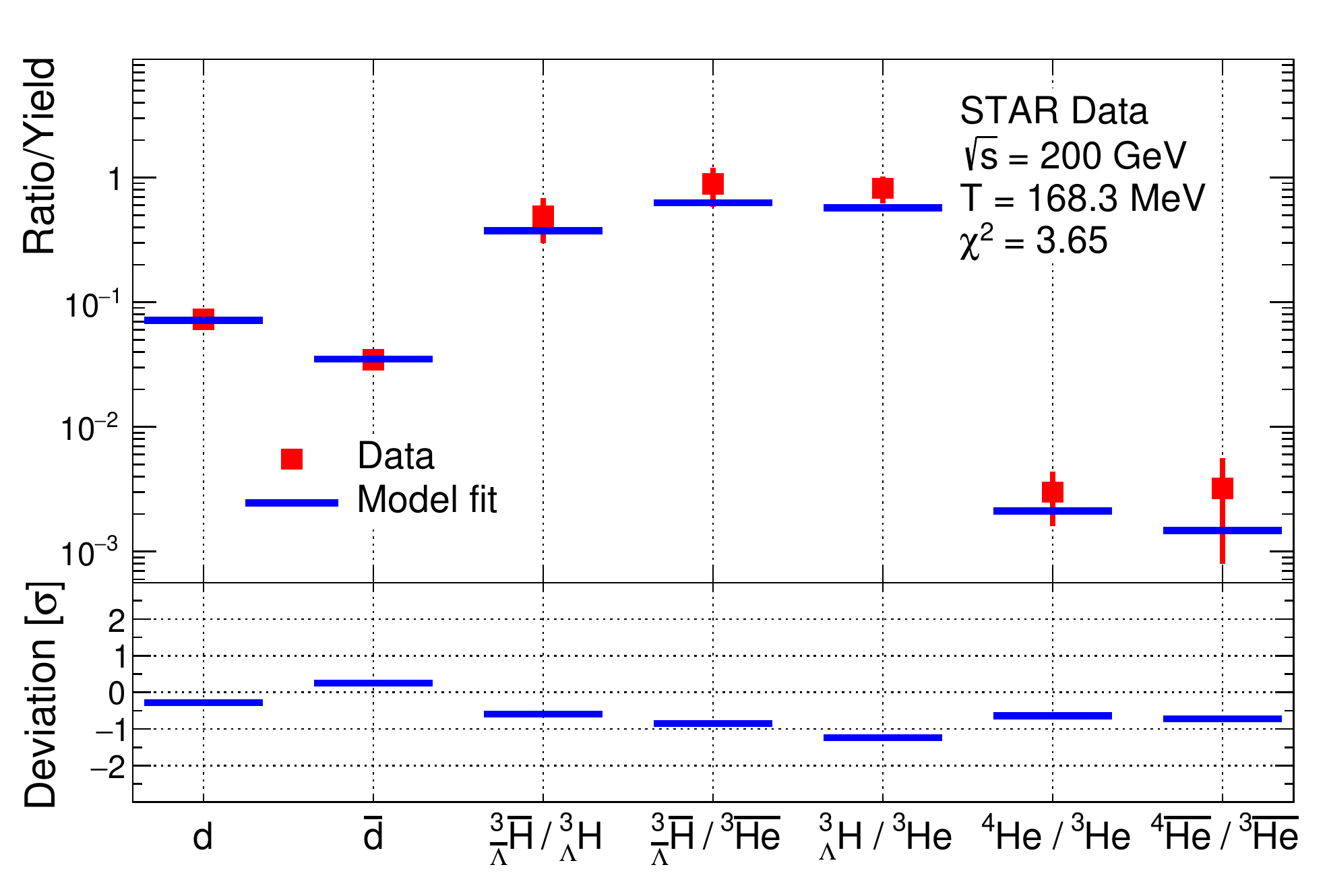}
	%\vspace*{-2mm}
	\caption{The yields of nuclear clusters measured at $\sqrt{s_{NN}} = 200$ GeV by STAR vs. theoretical description with IST$\Lambda$ EoS. Insertion shows the deviation of theory from data in	the units of experimental error.  
	{\bf Upper panel:} The $\min\chi^2_A(V)$ corresponds to the separate CFO of hadrons and nuclei.
	{\bf Lower panel:} Same as in the upper panel, but for the single CFO of hadrons an nuclei.}
	\label{STAR_fit}
\end{figure}

However, before going into a discussion of numerics we have to remind that  in the present work we follow the setup of Refs. \cite{HRGM_IST3,HRGM_IST6}, i.e. the hard-core radii of hadrons (new radii of Refs. \cite{HRGM_IST3}) given above,
the particle  table along with their decays and all the hadronic data of the  STAR Collaboration  measured at $\sqrt{s_{NN}} = 200$ GeV and the ones of the ALICE Collaboration measured at $\sqrt{s_{NN}} = 2.76$ TeV are taken from Refs. \cite{HRGM_IST3,HRGM_IST6}. Since all these elements of the MHRGM  are well documented, they can be found in the original works   \cite{HRGM_IST3,HRGM_IST6}, while here we concentrate on the new and essential features of the novel MHRGM worked out here. 

First we analyze the ALICE Collaboration data measured at $\sqrt{s_{NN}} = 2.76$ TeV.
From Fig. \ref{ALICE_chi2} and Table 1 one can see that the results on $\chi^2$ obtained with the BMR$\Lambda$ and IST$\Lambda$ EoS are very close to the ones which were obtained in \cite{HRGM_IST6}. 
An apparent reason for such a similarity is that all generic features of the present results coincide with the ones found in \cite{HRGM_IST6}.  

In the single CFO scenario we have  10 experimental data points for hadronic ratios, 8  yields of light nuclear clusters and 2 fitting parameters (CFO temperatures $T_h=T_A$ and volume $V$), while all chemical potentials are set to zero \cite{HRGM_IST3,HRGM_IST6}.
Similarly to  the results
of Ref. \cite{HRGM_IST6} we see  that the CFO temperatures  $T_h$ found by the BMR$\Lambda$ and IST$\Lambda$ EoS 
for the single CFO scenario are practically the same and, moreover, they   coincide with the ones found in \cite{HRGM_IST6}.  Indeed, from Table 1 one can see that for this scenario the CFO temperatures  $T_h \bigl|_{BMR\Lambda } = 150.29 \pm 1.92$ MeV  and  $T_h \bigl|_{IST\Lambda } = 150.39 \pm 1.90$ MeV differ from each other on 0.1 MeV which is negligible difference  compared to their errors.  

For the scenario of separate CFO of hadrons  and light nuclear clusters one obtains a very similar  behavior of $T_h$ and $T_A$ found by the  BMR$\Lambda$ and IST$\Lambda$ EoS, although the number of fitting parameters is 3, 
since $T_h \neq T_A$. 
Therefore, in contrast to our previous findings, the refined strategy based on Eqs.  (\ref{eq_12}) and (\ref{eq_16})   automatically provides  a very close location of the minima of  light nuclear clusters $\chi^2_A$  obtained by the  BMR$\Lambda$ and IST$\Lambda$ EoS. Consequently, with the refined  strategy one does not  need to search for the common description of the data by two models, since such a description is automatically located within the error bars of the found CFO parameters.  Nevertheless,  the BMR$\Lambda$ and IST$\Lambda$ EoS provide an additional cross-check of the  obtained results.  

 Moreover, the main conclusion of Ref.  \cite{HRGM_IST6} that at ALICE energy of collisions the separate CFO of light nuclei  is the most probable one,  is  confirmed by the more accurate MHRGM worked out here with the unprecedented accuracies   $\chi^2_{tot} / dof  \bigl|_{IST\Lambda } = 0.753$ and  $\chi^2_{tot} / dof  \bigl|_{BMR\Lambda } = 0.676$. 
 The main difference with Ref.  \cite{HRGM_IST6}   is that the CFO temperature of light nuclear clusters $T_A^{com} = 175.1^{+2.3}_{-3.9}$ MeV  \cite{HRGM_IST6}  decreased now  by a few MeV to the value $T_A  \bigl|_{IST\Lambda } = 169.25 \pm 5.57$ MeV  as one can see from Table 1 for the separate CFO scenario.  Apparently, these  values of  the CFO temperature of light nuclear clusters  overlap and   this means that there is no any inconsistency in the obtained results here and in Ref. \cite{HRGM_IST6}  and between the employed versions of EoS. 
 
 The results of  the ALICE data  fits  are shown in Fig.~\ref{ALICE_fit} for the scenario of separate CFO. From this figure one  can see that all experimental yields of light nuclear clusters   are described very well. Since the quality of the hadronic data description is the same as in \cite{HRGM_IST6}), these figures are not shown here.

\begin{figure}[ht]
	\centering
	\includegraphics[width=\columnwidth]{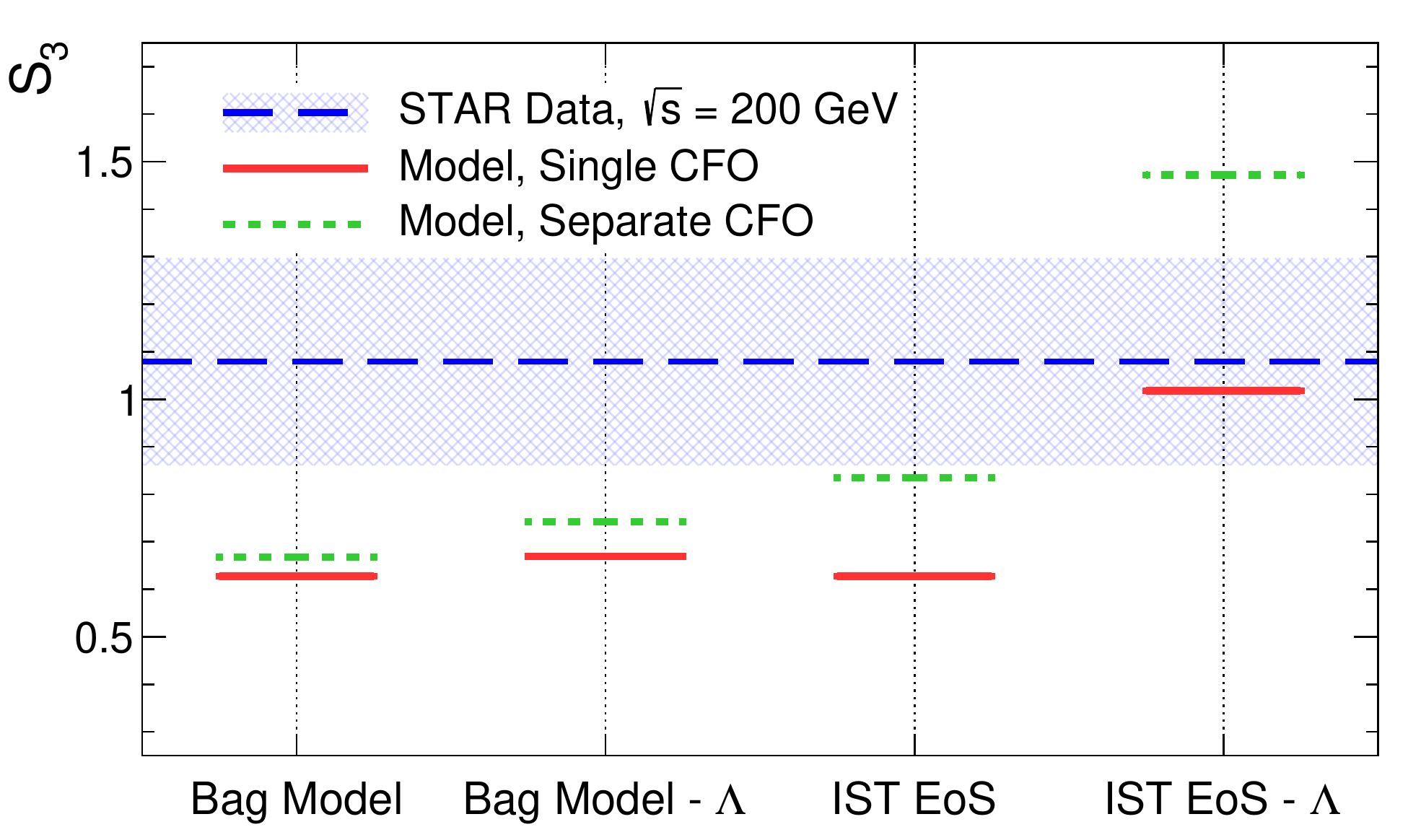}
	%\vspace*{-1mm}
	\includegraphics[width=\columnwidth]{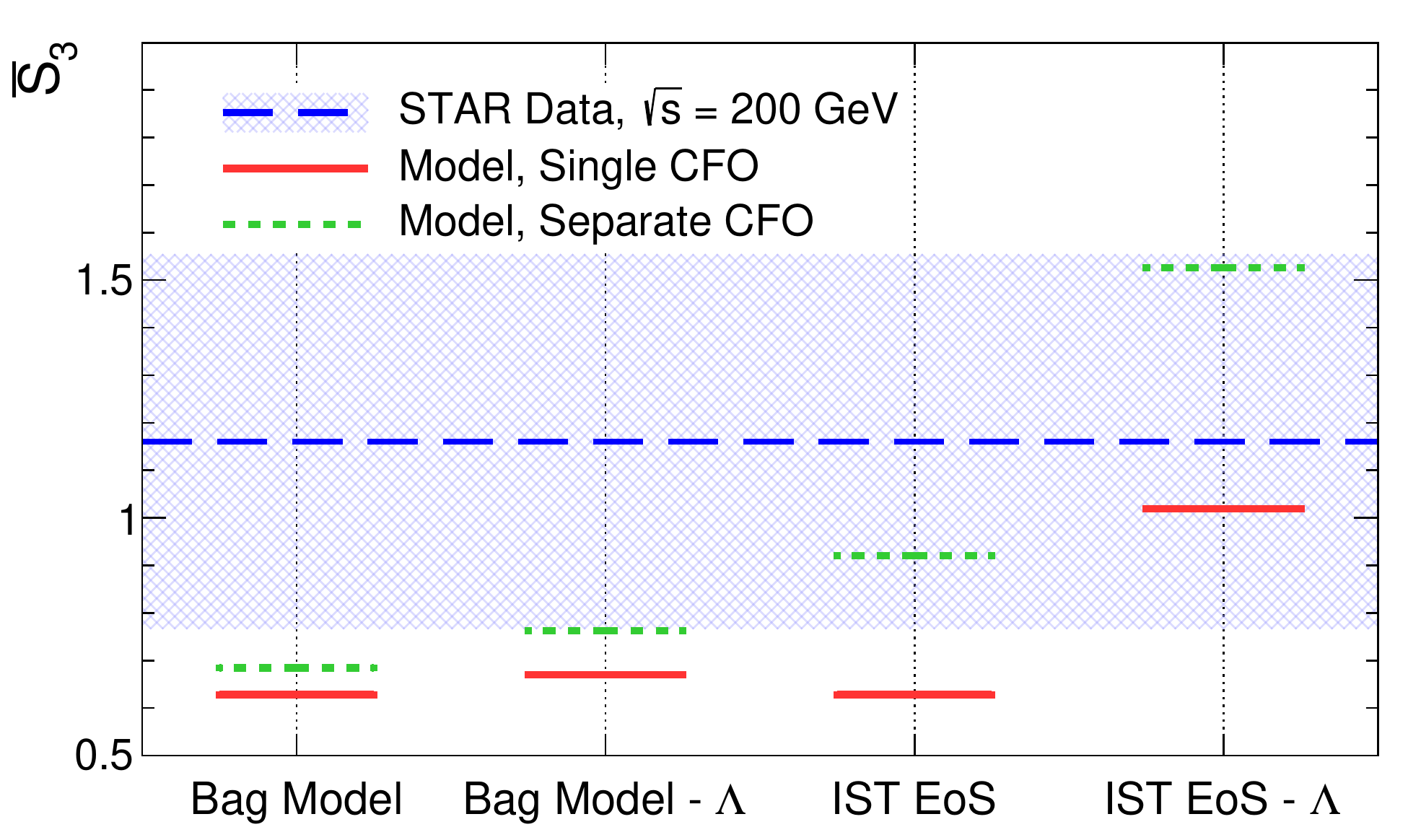}
	%\vspace*{-2mm}
	\caption{
	{\bf Upper panel:} $S_3$ ratio measured at $\sqrt{s_{NN}} = 200$ GeV by STAR  \cite{STARA1} vs. theoretical description obtained for  different CFO scenarios and EoS.
	{\bf Lower panel:} Same as in the upper panel, but for $\bar{S}_3$ ratio  \cite{STARA1}.}
	\label{STAR_S3}
\end{figure}

More interesting results are obtained for the light nuclear clusters  data measured by the STAR Collaboration at 
$\sqrt{s_{NN}} = 200$ GeV.  
These STAR data consist of 10 hadronic ratios (for more details see \cite{HRGM_IST6}), the (anti-)deu\-teron yields \cite{STARA3} and 5 other light nuclear clusters ratios \cite{STARA1,STARA2}.
They contain the PHTR ratios which no one could describe for almost a decade. 
From Table 2 one can see that the results obtained by the IST$\Lambda$ EoS for both CFO scenarios are practically the same and their fit quality is $\chi^2_{tot} / dof \simeq 1$, i.e. it is essentially reduced  compared to Ref.  \cite{HRGM_IST6}.

\begin{table*}[!ht]
	\label{chi2_ALICE_table}
	\centering
	\begin{tabular}[t]{lcccc}
		\toprule
		Description                  & $T_h,$  MeV      & $T_A,$  MeV       & $V_A,$ fm$^3$  & $\chi^2/dof$  \\ 
		\midrule
		
		Single CFO, IST$\Lambda$    & $150.29 \pm 1.92$ & $150.29 \pm 1.92$ & $13145 \pm 2233$ & 1.433  \\ 
		
		Single CFO, BMR$\Lambda$    & $150.39 \pm 1.90$ & $150.39 \pm 1.90$ & $11201 \pm 2009$ & 1.293  \\ 
		
		Separate CFO, IST$\Lambda$  & $148.12 \pm 2.03$ & $169.25 \pm 5.57$ & $3898 \pm 1272$ & 0.753  \\ 
		
		Separate CFO, BMR$\Lambda$  & $148.12 \pm 2.03$ & $167.59 \pm 5.39$ & $3123 \pm 1198$ & 0.676  \\ 
		
		\bottomrule
	\end{tabular}
	\caption{The results obtained by the advanced HRGM  for the fit of ALICE data measured at $\sqrt{s_{NN}} = 2.76$ TeV. The CFO temperature of hadrons is $T_h$, the CFO temperature of light (anti)nuclei is $T_A$, while their CFO volume is $V_A$. The last column gives the fit quality. } 
\end{table*}

\begin{table*}[ht]
	\label{chi2_STAR}
	\centering
	\begin{tabular}[t]{lcccccc}
		\toprule
		Description       & $T_h,$  MeV & $T_A,$ MeV & $\mu_{B}^{h},$ MeV & $\mu_{B}^{A},$ MeV & $V_A,$ fm$^3$ & $\chi^2/dof$  \\ 
		\midrule
		
		Single CFO, IST$\Lambda$   & $168.30 \pm 3.85$ & $168.30 \pm 3.85$ & $30.12 \pm 3.27$ & $30.12 \pm 3.27$ & $2056 \pm 375$ & 1.069  \\  
		
		Single CFO, BMR$\Lambda$   & $167.43 \pm 3.84$ & $167.43 \pm 3.84$ & $30.00 \pm 3.26$ & $30.00 \pm 3.26$ & $1667 \pm 355$ & 1.339  \\ 
		
		Separate CFO, IST$\Lambda$ & $166.51 \pm 4.07$ & $185.99 \pm 9.09$ & $28.84 \pm 5.37$ & $34.30 \pm 4.81$ & $1093 \pm 278$ & 0.995  \\ 
		
		Separate CFO, BMR$\Lambda$ & $166.51 \pm 4.07$ & $182.69 \pm 14.1$ & $28.84 \pm 5.37$ & $33.30 \pm 4.94$ & $831 \pm 455$ & 1.459  \\ 
		
		\bottomrule
	\end{tabular}
	\caption{The results obtained by the advanced HRGM  for the fit of STAR data measured at $\sqrt{s_{NN}} = 200$ GeV. The CFO temperature of hadrons (nuclei) is $T_h$ ($T_A$), the CFO baryonic chemical potential  of hadrons (nuclei) is $\mu_B^h$ ($\mu_B^A$), while the CFO volume of nuclei  is $V_A$. The last column gives the fit quality.} 
\end{table*}

From Fig. \ref{STAR_chi2} one can see that for  the separate CFO of light nuclear clusters  both EoS 
provide the CFO temperature of nuclei $T_A$ above 180 MeV (in this case there are 5  fitting parameters such as  the CFO temperature of hadrons (nuclei) $T_h$ ($T_A$), the CFO baryonic chemical potential  of hadrons (nuclei)  $\mu_B^h$ ($\mu_B^A$) and  the CFO volume of nuclei $V_A$).  Note, however,  that according to lattice version of  QCD  at 
vanishing value of the  baryonic chemical potential  \cite{KAB_lqcd1,KAB_lqcd2} 
 it is rather problematic to use the hadronic EoS for  these  CFO temperatures, since this is the region above the cross-over to quark-gluon plasma. 
Although  for the separate  CFO scenario  all the  PHTR  are reproduced by the  IST$\Lambda$ EoS with  the deviation smaller than 1$\sigma$ (see Fig. \ref{STAR_fit}),
this scenario can be ruled out by requiring a consistency with the lattice QCD results.

As an independent indicator in favor of the single  CFO scenario for the STAR energies the  $S_3$ and $\overline{S}_3$ ratios 
\begin{equation}\label{eq_18}
	S_3 = \frac{{}^{3}_{\Lambda}\text{H}}{{}^{3}\text{He} \times \frac{\Lambda}{p}}, \quad  \overline{S}_3 = \frac{{}^{3}_{\overline{\Lambda}}\overline{\text{H}}}{{}^{3}\overline{\text{He}} \times \frac{\overline{\Lambda}}{\overline{p}}},
\end{equation}  
can be used. 
From Fig. \ref{STAR_S3} one can clearly see that these special ratios provided by the STAR  Collaboration \cite{STARA1} are  well reproduced for the single CFO  scenario found by the IST$\Lambda$ EoS. 
%%%{\color{red} However,  a word of caution   has to be said here.  It can be that for the data may change with time.}
In our opinion  it is absolutely remarkable that the data on the  $S_3$ and $\overline{S}_3$ ratios, 
which were not used in our fits, are reproduced by the most elaborate version of the MHRGM for the single CFO scenario. 
In this case there are only 3 fitting parameters, namely the CFO temperature of hadrons and nuclei $T_h=T_A$, the CFO baryonic chemical potential  of hadrons and nuclei  $\mu_B^h = \mu_B^A$ and  the CFO volume of nuclei  $V_A$.

Using the CFO parameters determined  from the fits of STAR and ALICE data  discussed above, we made predictions for the  yields of some light nuclear clusters  which are also discussed in \cite{baltz} (see Tables 3 and 4)  with different CFO scenarios. As one can see   from Tables 3 and 4 the absolute yields of  the light nuclear clusters are somewhat  higher than those of the separate CFO scenario and, therefore,  the new measurements of some of these nuclei can help to distinguish the scenarios of CFO of nuclei. Furthermore, comparison of our predictions with the ones obtained for the same collision energies by the advanced coalescence model \cite{Sun2016} (see Table II therein)  shows that our numbers are systematically higher. Therefore,  the new measurements of  the yields of such nuclei as $_\Lambda^4 {H}$,  $_\Lambda^4{He}$ and $_{\Lambda \Lambda}^4{H}$ maybe of crucial importance  for reliable  determination of the CFO scenarios in the MHRGM and  in the coalescence model.

\begin{table*}[ht]
	\label{table_STAR}
	\centering
	\begin{tabular}[t]{lllllllll}
	\toprule 
	Particle  & BMR EoS & IST EoS & BMR$\Lambda$ EoS & IST$\Lambda$ EoS & BMR EoS & IST EoS & BMR$\Lambda$ EoS & IST$\Lambda$ EoS  \\ 
	$~$		  & (Single)  & (Single) & (Single) & (Single) & (Separate)  & (Separate) & (Separate) & (Separate)   \\ 
	\midrule
	
 ${}^{3}$H & $3.0\times 10^{-4}$ & $2.2\times 10^{-4}$ & $3.0\times 10^{-4}$ & $2.2\times 10^{-4}$ & $4.0\times 10^{-4}$ & $3.5\times 10^{-4}$ & $4.2\times 10^{-4}$ & $2.8\times 10^{-4}$ \\ 
${}^{3}$He & $3.0\times 10^{-4}$ & $2.2\times 10^{-4}$ & $3.0\times 10^{-4}$ & $2.2\times 10^{-4}$ & $4.0\times 10^{-4}$ & $3.5\times 10^{-4}$ & $4.2\times 10^{-4}$ & $2.8\times 10^{-4}$ \\ 
${}^{4}$He & $8.8\times 10^{-7}$ & $4.6\times 10^{-7}$ & $8.9\times 10^{-7}$ & $4.7\times 10^{-7}$ & $1.6\times 10^{-6}$ & $1.3\times 10^{-6}$ & $1.7\times 10^{-6}$ & $7.0\times 10^{-7}$ \\ 
${}_{\Lambda}^{3}$H & $1.0\times 10^{-4}$ & $7.7\times 10^{-5}$ & $1.1\times 10^{-4}$ & $1.3\times 10^{-4}$ & $1.5\times 10^{-4}$ & $1.6\times 10^{-4}$ & $1.7\times 10^{-4}$ & $2.3\times 10^{-4}$ \\ 
${}_{\Lambda}^{4}$H & $1.1\times 10^{-6}$ & $5.9\times 10^{-7}$ & $1.0\times 10^{-6}$ & $1.0\times 10^{-6}$ & $2.1\times 10^{-6}$ & $2.3\times 10^{-6}$ & $2.1\times 10^{-6}$ & $2.2\times 10^{-6}$ \\ 
${}_{\Lambda}^{4}$He & $1.1\times 10^{-6}$ & $5.9\times 10^{-7}$ & $1.0\times 10^{-6}$ & $1.0\times 10^{-6}$ & $2.1\times 10^{-6}$ & $2.3\times 10^{-6}$ & $2.1\times 10^{-6}$ & $2.2\times 10^{-6}$ \\ 
${}_{\Lambda}^{5}$He & $3.1\times 10^{-9}$ & $1.1\times 10^{-9}$ & $2.4\times 10^{-9}$ & $1.9\times 10^{-9}$ & $7.7\times 10^{-9}$ & $1.1\times 10^{-8}$ & $6.6\times 10^{-9}$ & $4.8\times 10^{-9}$ \\ 
${}_{\Lambda\Lambda}^{~~4}$H & $2.8\times 10^{-7}$ & $1.5\times 10^{-7}$ & $3.6\times 10^{-7}$ & $4.1\times 10^{-7}$ & $5.7\times 10^{-7}$ & $7.7\times 10^{-7}$ & $9.2\times 10^{-7}$ & $1.3\times 10^{-6}$ \\ 
${}_{\Lambda\Lambda}^{~~5}$H & $9.0\times 10^{-10}$ & $3.1\times 10^{-10}$ & $9.8\times 10^{-10}$ & $9.4\times 10^{-10}$ & $2.4\times 10^{-9}$ & $4.4\times 10^{-9}$ & $3.3\times 10^{-9}$ & $3.6\times 10^{-9}$ \\ 
${}_{\Lambda\Lambda}^{~~5}$He & $9.0\times 10^{-10}$ & $3.2\times 10^{-10}$ & $9.9\times 10^{-10}$ & $9.5\times 10^{-10}$ & $2.4\times 10^{-9}$ & $4.4\times 10^{-9}$ & $3.3\times 10^{-9}$ & $3.6\times 10^{-9}$ \\ 
${}_{\Lambda\Lambda}^{~~6}$He & $2.1\times 10^{-12}$ & $4.5\times 10^{-13}$ & $1.9\times 10^{-12}$ & $1.5\times 10^{-12}$ & $7.5\times 10^{-12}$ & $2.4\times 10^{-11}$ & $8.6\times 10^{-12}$ & $6.2\times 10^{-12}$ \\ 
%$(\Lambda n)_b$ & $2.6\times 10^{-2}$ & $2.6\times 10^{-2}$ & $3.3\times 10^{-2}$ & $4.2\times 10^{-2}$ & $2.8\times 10^{-2}$ & $3.5\times 10^{-2}$ & $3.9\times 10^{-2}$ & $6.0\times 10^{-2}$ \\ 
%$(\Lambda\Lambda)_b$ & $3.3\times 10^{-3}$ & $3.3\times 10^{-3}$ & $6.4\times 10^{-3}$ & $8.5\times 10^{-3}$ & $3.7\times 10^{-3}$ & $5.7\times 10^{-3}$ & $9.7\times 10^{-3}$ & $1.7\times 10^{-2}$ \\ 

	\bottomrule
\end{tabular}
	\caption{Expected yields of some (hyper-)nuclei per event obtained by the advanced HRGM for the fit of STAR data measured at $\sqrt{s_{NN}} = 200$ GeV with different CFO scenarios and nuclear cluster radii estimations. The CFO scenario is specified in parentheses.} 
\end{table*}

\section{Conclusions and perspectives} \label{Conclusions}

In this work we obtained an unprecedentedly accurate description of the hadronic and light nuclear clusters data measured
by the STAR Collaboration at  $\sqrt{s_{NN}} =200$ GeV and by  the ALICE LHC at 
$\sqrt{s_{NN}} =2.76$ TeV with $\chi^2/dof  \simeq \frac{26.261}{18-3+17-3} \simeq 0.91$ for two best fits of combined data. This  is achieved by applying the new strategy of analyzing the light nuclear clusters data and by using the correct value of the hard-core radius of the (anti-)$\Lambda$ hyperons in the expressions for the classical second virial coefficients and for the 
equivalent   hard-core radius of HTR. The generalized expressions for these quantities are worked out here.  Using them we obtained a greatly improved description of the STAR data compared to our recent analysis done in  \cite{HRGM_IST6}. At the same time  for the ALICE data we obtained  almost the same results and the same high quality of the fit. 

The fact, that the hard-core radius of the (anti-)$\Lambda$ hyperons $R_\Lambda$ is essentially smaller than the one of other hadrons was for the first time justified in Ref. \cite{Sagun14}. The high quality description of the hadronic multiplicities 
measured at the collision energies  $\sqrt{s_{NN}} =2.7, 3.3, 3.8, 4.3, 4.9,$ $ 6.3, 7.7, 8.8,
 9.2, 12.3, 17.3,  62.4, 130,  200, 2760$ GeV  obtai\-ned  by the most  elaborate version of  MHRGM with the IST  EoS for the same set of hadronic hard-core radii  
\cite{HRGM_IST2,HRGM_IST3} confirmed this conclusion about the hard-core radius  of the (anti-)$\Lambda$ hyperons.
Finally, the resolution of the PHTR puzzle provides a strong support for the model assumption that $R_\Lambda \simeq 0.085$ fm is much smaller
than  $R_b \simeq 0.365$ fm. 
This is in accord with the expectation from Pauli blocking phenomenology, because nucleons and hyperons have a different quark content so that the repulsive force induced by quark exchange effects between them is much smaller than between nucleons.

One of the most striking findings of this work is that for  the most probable scenario of CFO for the STAR energy, i.e. for a single CFO of hadrons and nuclei,  the obtained parameters allowed us to predict the ratios $S_3$ and $\overline{S}_3$  \cite{STARA1} in accordance with the experimental
data  without fitting them 
at all. 
Apparently, this 
fact evidences about  an internal consistency of the approach based on the  IST$\Lambda$ EoS and about  a high reliability of the  obtained results. Based on these findings   we predicted   the absolute yields  of several  light nuclear clusters including  some exotic ones for the considered energies of collisions. 

In contrast to all  previous findings,  the CFO temperature $T_A \simeq  168.75 \pm 5.57$ MeV  of the most probable scenarios of light nuclear clusters CFO  is  found  to be the same  for the STAR and ALICE data. 
Above it was argued  that this is the hadronization temperature of  quark-gluon bags. 
Note that this conclusion is in line with the lattice QCD results for the chiral cross-over transition region which at vanishing baryon chemical potential \cite{KAB_lqcd1,KAB_lqcd2} has a full-width at half maximum of about 40 MeV of the continuum-extrapolated chiral susceptibility which peaks at the 
pseudocritical transition temperature $T_c=156.5\pm 1.5$ MeV \cite{Bazavov:2018mes}.

It is necessary to stress that the MHRGM worked out during last few years  in \cite{HRGM_IST1,HRGM_IST2,HRGM_IST3,HRGM_IST4,HRGM_IST5,HRGM_IST6} 
creates  the new and rather high standards  in the description of the experimental multiplicities of hadrons 
and light nuclear clusters. 
Moreover, this work  proves once again that   MHRGM based on the IST concept,
is the most efficient, reliable  and convenient tool  to  elucidate the subtleties of the CFO process in high energy nuclear
collisions. 
%{\color{red}
Therefore,  the question arises whether the successful phenomenology of the MHRGM based on the IST concept 
could find theoretical support by the microscopic  approaches like the one suggested recently  \cite{Blaschke:2020} 
in particular about the apparent independence of the hard core radii on temperature and density of the medium 
accessible in heavy ion collision experiments
and about their dependence on the quark content of the hadrons involved.
%}
Furthermore,  when in the future experiments at NICA JINR and FAIR GSI the accuracy of the collected  
data will be increased by an order of magnitude, the  MHRGM based on the IST concept will be the key 
element of the phenomenological analysis because of its advantages over the other versions of the HRGM. 

It is clear that,  similarly to the usual chemistry,  sooner or later the task of  hadrochemistry  
to  determine and to  tabulate  the classical second virial coefficients of all measured  hadrons
will be on demand.
The IST EoS and its generalizations,
which account  for the curvature tension  term \cite{IST4,IST5}, are perfectly suited for this task,
since they allow one to model the high virial coefficients of classical and quantum systems with very high
accuracy.  

Furthermore, the examples of different mixtures studied in Ref. \cite{IST5}  by the EoS with the  induced surface and curvature tensions   give us  a good hope that such EoS can be used  to reliably model the mixtures of  hadrons, nuclei and QGP bags,  to elucidate  the subtleties of interaction between the hadrons, nuclei and QGP bags
and to study the phase transformations among them that occur  in high energy nuclear collisions.\\

\noindent
{\bf Author contributions.}
%{\color{blue}
O.V.V. suggested  the main idea behind this work and   together with K.A.B.  he developed a novel  formalism 
summarized in Eqs.  (10)-(17). 
K.A.B.  together with D.B.B. took the lead in writing the manuscript.
O.V.V. and E.S.Z. performed fit of the experimental data on the light (anti)nuclei and hadrons under the supervision of L.V.B., and E.E.Z..
O.V.V., E.S.Z.,  L.V.B. and E.E.Z.   designed the figures. G.M.Z., L.V.B., E.E.Z.  contributed to the interpretation of the results and provided a critical feedback. All authors discussed the results and contributed to the final version of manuscript.\\
%}

	\noindent
	{\bf Acknowledgments.} The authors are thankful to Boris Grinyuk,  Dmytro Oliinychenko and Ivan Yakimenko   for  the valuable comments. 
K.A.B.  and G.M.Z. acknowledge support from the NAS  of Ukraine by its priority project ``Fundamental properties of the matter in the relativistic collisions of nuclei and in the early Universe"
(No. 0120U100935).
The work of L.V.B. and E.E.Z. was supported by the Norwegian Research Council (NFR) under grant No. 255253/F53 CERN Heavy Ion Theory, and by the 
Russian Fund for Basic Research (RFBR) grants No. 18-02-40085 and No. 18-02-40084. 
 K.A.B., O.V.V.,   and  L.V.B.  thank the Norwegian Agency for International Cooperation and $ $ Quality Enhancement in Higher Education for the financial support under grants CPEA-LT-2016/10094 and UTF-2016-long-term/10076. $ $ 
D.B.B. acknowledges partial support from RFBR under grant No. 18-02-40137 and from the National Research Nuclear University (MEPhI) in the framework of the Russian Academic Excellence Project under contract No. 02.a03.21.0005.  
The authors  are  grateful to the COST Action CA15213 ``THOR" for supporting their  networking. 

\begin{table*}[ht]
	\label{table_ALICE}
	\centering
	\begin{tabular}[t]{lllllllll}
	\toprule 
	Particle  & BMR EoS & IST EoS & BMR$\Lambda$ EoS & IST$\Lambda$ EoS & BMR EoS & IST EoS & BMR$\Lambda$ EoS & IST$\Lambda$ EoS  \\ 
	$~$		  & (Single)  & (Single) & (Single) & (Single) & (Separate)  & (Separate) & (Separate) & (Separate)   \\ 
	\midrule
		
 ${}^{3}$H & $2.3\times 10^{-4}$ & $1.9\times 10^{-4}$ & $2.3\times 10^{-4}$ & $2.0\times 10^{-4}$ & $3.4\times 10^{-4}$ & $3.3\times 10^{-4}$ & $3.4\times 10^{-4}$ & $2.6\times 10^{-4}$ \\ 
${}_{\Lambda}^{4}$H & $4.0\times 10^{-7}$ & $2.7\times 10^{-7}$ & $3.8\times 10^{-7}$ & $3.6\times 10^{-7}$ & $1.2\times 10^{-6}$ & $1.2\times 10^{-6}$ & $1.0\times 10^{-6}$ & $1.1\times 10^{-6}$ \\ 
${}_{\Lambda}^{4}$He & $4.0\times 10^{-7}$ & $2.8\times 10^{-7}$ & $3.8\times 10^{-7}$ & $3.7\times 10^{-7}$ & $1.2\times 10^{-6}$ & $1.2\times 10^{-6}$ & $1.0\times 10^{-6}$ & $1.1\times 10^{-6}$ \\ 
${}_{\Lambda}^{5}$He & $5.3\times 10^{-10}$ & $3.0\times 10^{-10}$ & $4.6\times 10^{-10}$ & $4.0\times 10^{-10}$ & $2.8\times 10^{-9}$ & $1.9\times 10^{-9}$ & $2.1\times 10^{-9}$ & $1.7\times 10^{-9}$ \\ 
${}_{\Lambda\Lambda}^{~~4}$H & $9.2\times 10^{-8}$ & $6.3\times 10^{-8}$ & $1.1\times 10^{-7}$ & $1.1\times 10^{-7}$ & $3.1\times 10^{-7}$ & $3.6\times 10^{-7}$ & $3.8\times 10^{-7}$ & $4.8\times 10^{-7}$ \\ 
${}_{\Lambda\Lambda}^{~~5}$H & $1.4\times 10^{-10}$ & $7.8\times 10^{-11}$ & $1.5\times 10^{-10}$ & $1.4\times 10^{-10}$ & $8.6\times 10^{-10}$ & $7.0\times 10^{-10}$ & $8.8\times 10^{-10}$ & $9.1\times 10^{-10}$ \\ 
${}_{\Lambda\Lambda}^{~~5}$He & $1.4\times 10^{-10}$ & $7.9\times 10^{-11}$ & $1.5\times 10^{-10}$ & $1.4\times 10^{-10}$ & $8.6\times 10^{-10}$ & $7.1\times 10^{-10}$ & $8.9\times 10^{-10}$ & $9.2\times 10^{-10}$ \\ 
${}_{\Lambda\Lambda}^{~~6}$He & $1.6\times 10^{-13}$ & $6.9\times 10^{-14}$ & $1.5\times 10^{-13}$ & $1.3\times 10^{-13}$ & $1.7\times 10^{-12}$ & $5.5\times 10^{-13}$ & $1.5\times 10^{-12}$ & $1.2\times 10^{-12}$ \\ 
%$(\Lambda n)_b$ & $3.8\times 10^{-2}$ & $3.8\times 10^{-2}$ & $4.4\times 10^{-2}$ & $5.0\times 10^{-2}$ & $3.7\times 10^{-2}$ & $4.3\times 10^{-2}$ & $4.6\times 10^{-2}$ & $6.2\times 10^{-2}$ \\ 
%$(\Lambda\Lambda)_b$ & $4.5\times 10^{-3}$ & $4.5\times 10^{-3}$ & $6.6\times 10^{-3}$ & $7.6\times 10^{-3}$ & $4.9\times 10^{-3}$ & $6.7\times 10^{-3}$ & $9.4\times 10^{-3}$ & $1.3\times 10^{-2}$ \\   		
	\bottomrule
	\end{tabular}
	\caption{Expected yields of some (hyper-)nuclei per event obtained by the advanced HRGM for the fit of ALICE data measured at $\sqrt{s_{NN}} = 2.76$ TeV with different CFO scenarios. The CFO scenario is specified in parentheses.} 
\end{table*}

\end{document}